\documentclass[12pt,preprint]{aastex}
\usepackage{CJK}
\shortauthors{} 
\shorttitle{6$\mu$m}

\received{}
\usepackage{graphicx}
\graphicspath{{}}
\begin{document}

\title{The 6 $\mu$m Feature as A Tracer of Aliphatic Components of Interstellar Carbonaceous Grains}

\begin{CJK*}{Bg5}{bsmi}
\CJKtilde

\author{Chih-Hao Hsia (®L§Ó¯E), Seyedabdolreza Sadjadi, Yong Zhang (±iªa), Sun Kwok (³¢·s)\altaffilmark{1}}
\affil{Laboratory for Space Research, Faculty of Science, The University of Hong Kong, Pokfulam Road, Hong Kong, China}
\email{xiazh@hku.hk; ssadjadi@hku.hk; zhangy96@hku.hk; sunkwok@hku.hk}

\altaffiltext{1}{Visiting Professor, Department of Physics and Astronomy, University of British Columbia, Vancouver, B.C., Canada}

\begin{abstract}
An unidentified infrared emission (UIE) feature at 6.0 $\mu$m is detected in a number of astronomical sources showing the UIE bands.  In contrast to the previous suggestion that this band is due to C=O vibrational modes, we suggest that the 6.0 $\mu$m feature arises from olefinic double-bond functional groups.  These groups are likely to be attached to aromatic rings which are responsible for the major UIE bands.  The possibility that the formation of these functional groups is related to the hydrogenation process is discussed.            
\end{abstract}


\section{Introduction}

A family of strong unidentified infrared emission (UIE) bands at 3.3, 6.2, 7.7--7.9, 8.6, 11.3 and 12.7\,$\mu$m 
are commonly observed in astronomical sources  \citep[for a recent review, see][]{pe13}. 
As the UIE bands are ubiquitous  in  circumstellar envelopes, molecular clouds, and active galaxies, the nature of their carrier provides key information
for understanding the matter cycle and chemical evolution of galaxies.  Although the exact chemical structure of the carrier of the UIE bands are  still uncertain \citep{cataldo2013, papoular2013, jones2013, Kwok13,zk15}, it is generally accepted that the bands arise from C$-$H/C$-$C stretching and bending vibrational modes of aromatic and/or aliphatic compounds. Nevertheless, there also exists a small group of weak UIE bands that do not alway appear together with the stronger UIE bands.  Their nature is completely unknown. A valid assignment of their vibrational modes of these weaker bands would help to  identify the UIE carrier.   The 6.0\,$\mu$m feature is one of the members of this group.

Although the 6.0\,$\mu$m feature has been detected in various objects showing UIE bands \citep[e.g.,][]{Peeters02}, there has not yet been a thorough investigation of the nature of this feature.  The spectra of young stellar objects usually exhibit a strong absorption feature at 6.0\,$\mu$m, which is commonly ascribed to the bending mode of amorphous water ice \citep{ta87}.  However, solid H$_2$O  has a stronger absorption at 3.0\,$\mu$m, which has never been observed in these UIE sources. Furthermore, given the carbon-rich nature of UIE sources, H$_2$O can be ruled out as the carrier of the 6.0\,$\mu$m feature.

An early suggestion for the origin of the 6.0 $\mu$m  feature is a C$-$C stretching mode in a minor population of PAH molecules \citep{Beintema96}.   The position of the 6.0\,$\mu$m feature is close to that of the stretching mode of C=O and it has been suggested that HCOOH can be  a cause for broadening of the observed 6.0\,$\mu$m absorption feature in young stellar objects \citep{st96}.   This has led to the suggestion that the 6.0\,$\mu$m emission band in UIE sources is a carbonyl stretching mode from oxygenated polycyclic aromatic hydrocarbons (PAHs)  \citep{Peeters02}.  However, the C=O stretching mode lies in a wavelength shorter than 6.0\,$\mu$m so the identification is not perfect.  Another possibility is that the 6.0\,$\mu$m UIE arises from heteroatomic aromatic compounds. For example, \citet{js09} shows that the C-C mode with C atoms bounded to Si atoms can shift to the 6.0\,$\mu$m  range, and this feature  might suggest the presence of multiple Si-atom complexes.

In this paper, we attempt to understand the  origin of the 6.0\,$\mu$m UIE band through comparisons between the observations, laboratory measurements, and the theoretical computations resulting from quantum-chemical calculations.

\section{Observations and Data reduction} 


We present the analysis of 20 sources with 6.0 $\mu$m feature that include H{\sc ii}  or compact H{\sc ii} regions (CHIIs), Herbig Ae/Be (HAe/Be) stars, young planetary nebulae (YPNe), post-asymptotic giant branch stars (PAGBs), emission-line star (Em star),  R Coronae Borealis stars (RCBs), K giants, and galaxies.  These sources are selected from \citet{Peeters02}, \citet{Hrivnak00}, \citet{Clayton11}, \citet{de15} and \citet{Garcia11a}.

The spectra of these sources were extracted from the {\it ISO} and {\it Spitzer} database. These spectra were examined for possible presence of the 6.0 $\mu$m feature.  
The nature of individual objects are listed in column 10 of Table 1.  Each object is also classified as carbon-rich (C) or of mixed chemistry (M) based on features in the infrared spectra.  These classifications are listed in column 5 of Table 1.

\subsection{ISO SWS Observations}

The mid-infrared spectra of 13 objects in our sample are obtained with  the Short-Wavelength Spectrometer (SWS) on board the {\it Infrared Space Observatory (ISO)}.  The observations were performed using the Astronomical Observation Template (AOT) 01 mode at various speeds with resolving power ($\lambda$/$\triangle\lambda$) ranging from 500 to 1600, covering a wavelength range from 2.4--45.2 $\micron$. The diaphragm size is 14$\arcsec\times$ 20$\arcsec$ at 6 $\mu$m in SWS module and the on-source integration times of the measurements were between 1061 s and 6546 s, depending on the source brightnesses. The observations for all objects are centered on the core positions. 

The data reduction and calibration were performed through the SWS Interactive Analysis (IA) system \citep{Wieprecht98} and {\it ISO Spectral Analysis Package} (ISAP; Sturm et al. 1998). Dark current subtraction, scan direction matching, and absolute responsivity correction were done, interactively. To improve the signal-to-noise ratio (S/N) of {\it ISO} observations, the final SWS spectra were reduced using the combined data. The journal of {\it ISO} spectroscopic observations is summarized in Table \ref{tab1}.    

\subsection{Spitzer IRS spectra}

Three K-giant stars (HD 233517, PDS 100, and PDS 365), two RCBs (DY Cen and HV 2671) and two PAGBs (IRAS 14316-3920 and IRAS 14429-4539) were observed between 2004 March and 2009 April using {\it Spitzer} {\it Infrared Spectrograph} \citep[IRS;][]{houck04}. All of these objects were observed using the Short-Low (SL) module (5.2--14.5 $\micron$) with spectral dispersion of $R\sim$ 64--128. The diaphragm size is 3$\farcs$6 $\times$ 57$\arcsec$ in SL module and the total integration times of IRS observations range from 13 s to 258 s, depending on the sources' expected mid-infrared fluxes. A summary of IRS observations is also given in Table \ref{tab1}.

Data were reduced starting with basic calibrated data from the Spitzer Science Center's pipeline version s18.7 and were run through the {\bf IRSCLEAN} program to remove rogue pixels. Next we employed point-source spectral extractions to extract spectra using the Spectral Modeling, Analysis and Reduction Tool  \citep[SMART;][]{higdon04}. A final spectrum was produced using the combined IRS observations to improve the S/N.

\subsection{Extended source calibration}

The flux calibrations of the spectra in our sample are derived assuming that all objects are point sources for both {\it ISO} SWS and {\it Spitzer} IRS observations.   From the source sizes listed in Table 1, we can see that there are several  extended sources (two CHIIs and three galaxies) and extended source flux corrections may be needed for these objects.
However,  since both the 6.0 and 6.2 $\mu$m features were observed in the  same AOT band and  with the same aperture size, the extended source flux corrections will apply equally to both features.  For the purpose of determining the ratios of the 6.0 and 6.2 $\mu$m features and the  peak positions of the bands, our results will not be affected.

The uncertainties in fluxes are obtained from the noise level in the continuum. If we take into account the uncertainties of absolute flux calibrations and band flux measurements, the flux errors at these bands are estimated to be about 11--20\% and 13--25 \% for {\it ISO SWS} and {\it Spitzer IRS} observations, respectively.

\section{Observational results}
\label{results}

\begin{figure*}
\begin{center}
\begin{tabular}{c}
\resizebox{140mm}{!}{\includegraphics{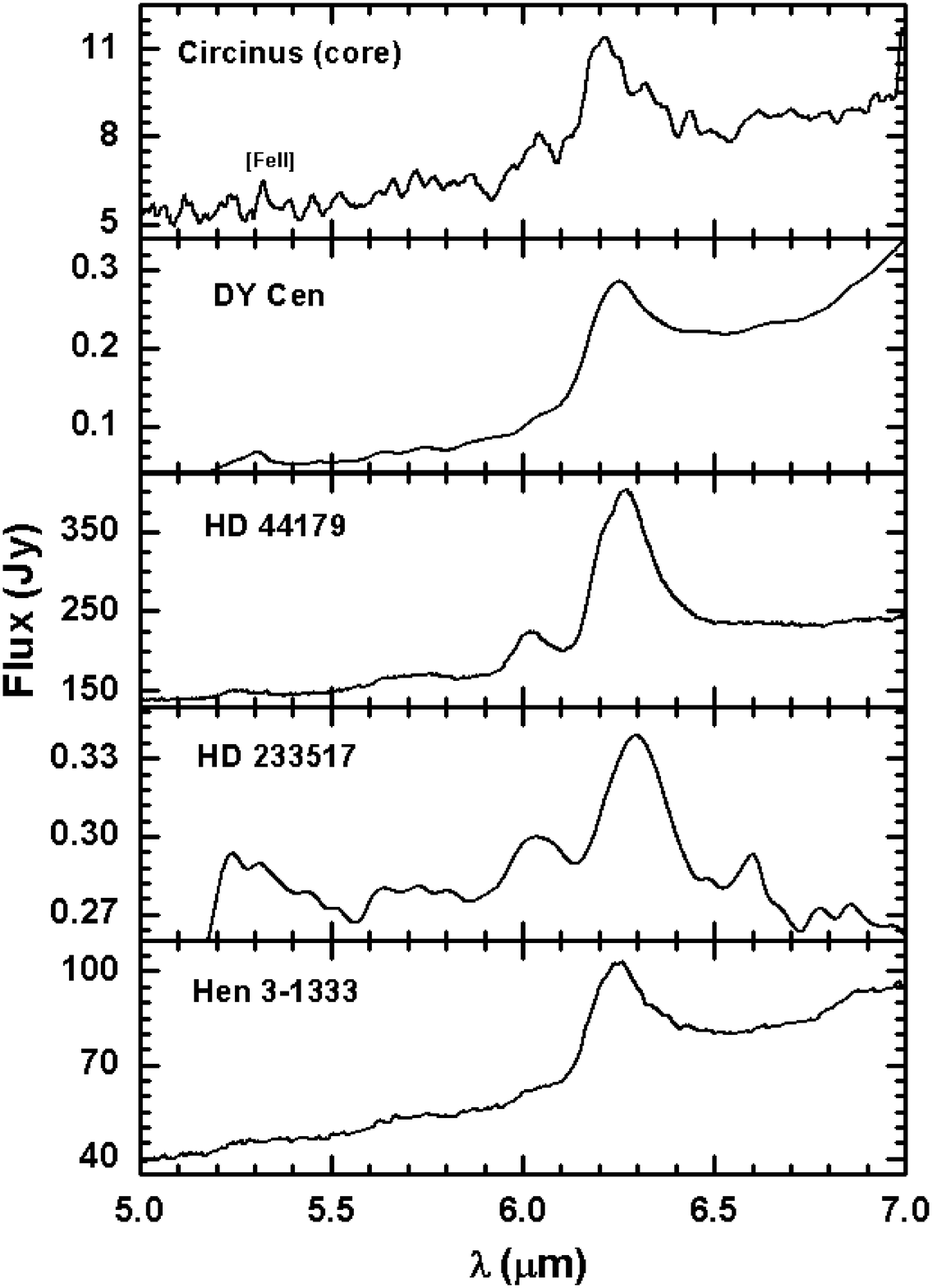}} \\
\end{tabular}
\end{center}
\caption{{\it ISO} and {\it Spitzer} 5--7 $\micron$ spectra of twenty objects. The most prominent feature in this spectral region is the 6.2 $\micron$ feature.  The [\ion{Fe}{2}] fine-structure line at 5.34 $\micron$ line can be seen in some objects. A weak feature at 6.0 $\micron$ is obviously present in some objects.}
\label{spectra}
\end{figure*}

Figure \ref{spectra} shows the  SWS and IRS spectra in the 5--7 $\mu$m region of the 20 objects in our sample.  A strong continuum can be seen in all cases.  The most prominent feature above the continuum is the UIE band around 6.2--6.3 $\micron$, which is widely observed as part of the UIE band family in many carbon-rich environments \citep{Molster96, Verstraete96, Jura06, perea09, Stanghellini12}.   The typical profile of the 6.2 $\micron$ emission feature shows a steep blue side and a long red tail. There are also some variations in the position of the peak, which has been attributed to variation in the aromatic to aliphatic ratio  \citep{sloan2007, Pino08}. 
A closer inspection of Figure \ref{spectra} also shows a weak feature centered near 6.0 $\micron$.   
The two narrow features in M82 and NGC 253   are the fine-structure line of [\ion{Fe}{2}] at 5.34 $\micron$ and H$_{2}$ S(5) molecular line at 6.91 $\micron$.

In order to better show the intrinsic strengths and profiles of the emission features, we have subtracted the underlying continua from the observed spectra.
The continuum fitting is performed using a third-order polynomial function  splined through the points in the feature-free spectral regions.  The feature-free regions provide adequate baselines for the fitting and we believe the uncertainty on the feature profile introduced by the continuum subtraction is small.
In five UIE sources (Circinus, M 82, S87 IRS1, W28 A2, and WL 16), extinction due to interstellar silicate absorption feature at 9.7 $\micron$  may also affect the continuum subtraction \citep{Spoon02}; but in the 5--7 $\mu$m region, this effect should be minimal.

\subsection{Central wavelengths and line widths}

\begin{figure*}
\begin{center}
\begin{tabular}{c}
\resizebox{140mm}{!}{\includegraphics{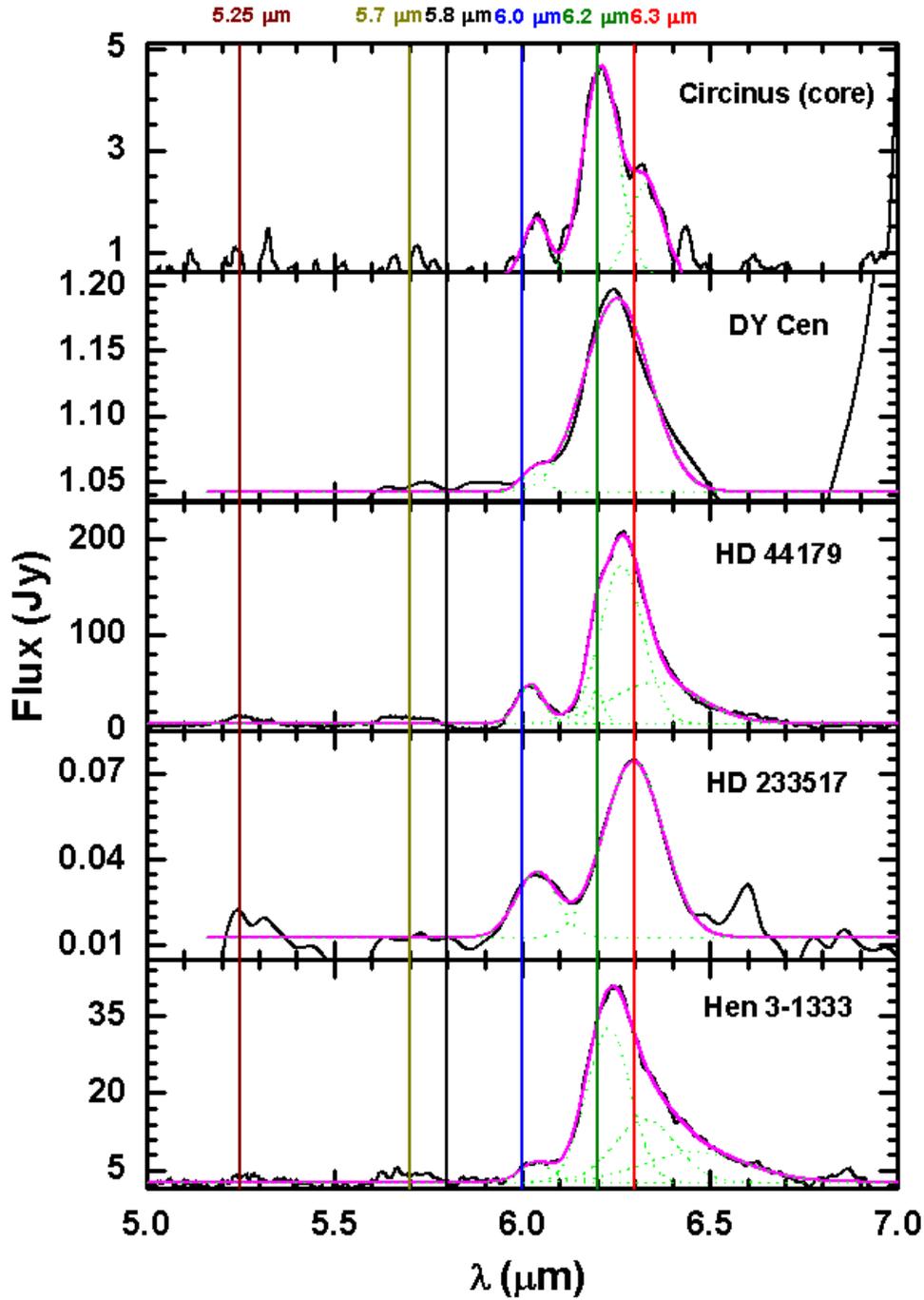}} \\
\end{tabular}
\end{center}
\caption{Continuum-subtracted 5--7 $\mu$m spectra of the twenty sample objects in Fig.~\ref{spectra}. The continua are fitted using 
third-order polynomials. The wine, dark yellow, blue, dark green, and red solid lines indicate the positions of features at 5.25, 5.7, 6.0, 
6.2, and 6.3 $\micron$, respectively. We do not see the carbonyl $>$C=O stretching vibrational band at 5.8 $\micron$ (black solid lines). 
The light green dotted lines are decomposed components of 6.0 and 6.2 $\micron$ features and the pink line is the sum of all band profiles.
}
\label{subtracted}
\end{figure*}

The continuum-subtracted spectra are shown in Figure \ref{subtracted}. The strengths and central wavelength ($\lambda_{c}$) of the 6.0 $\micron$ feature can be seen to vary  from source to source.
We can also see a weak UIE band around 6.9 $\micron$ in the spectra of IRAS 17347$-$3139. This feature is most likely due to bending modes of aliphatic side groups (e.g., methylene $-$CH$_{2}-$) \citep{Sandford13}. In addition, two weak emission features at 5.25 $\micron$ and 5.7 $\micron$ can be seen in several sources (e.g., HD 44179, HD 233517, Hen 3-1333, M 82, PDS 365, and S87 IRS1).  These two features are a blend of combination and overtone bands involving CC stretching, CH in-plane and CH out-of-plane bending fundamental vibrations of aromatic compounds  \citep{Boersma09, Roche96}. We note that no carbonyl $>$C=O stretching vibrational band at 5.8 $\micron$ is detected in the continuum-subtracted spectra of these objects. 

The peak wavelengths of the 6.0 and  6.2 $\micron$ features as measured from continuum-subtracted spectra (Figure \ref{subtracted}) of these objects are listed in Table \ref{tab2}.  The names of sources are listed in Column 1. Columns 2 and 3 give the observed 6.0 $\micron$ feature central wavelength ($\lambda_{c}$) and Full-Width Half-Maximum (FWHM) of the feature, respectively.  Column 4 gives the peak wavelength of the 6.2 $\mu$m feature.  For the 6.0 $\mu$m band, all values are measured using a Gaussian fitting routine.   
The peak wavelength of the 6.0 $\micron$ feature varies from 6.000 to 6.065 $\micron$, with an average value of 6.031 $\pm$ 0.020 $\micron$. The average FWHM of these objects is 0.081$\pm$0.022 $\mu$m for the 6.0 $\mu$m feature.
Unlike the 6.2 $\mu$m feature, the profiles of the 6.0 $\mu$m features are not obviously asymmetric.

Since the 6.2 $\mu$m feature has an asymmetric profile, we have tried to fit the profiles of this feature  with three Gaussians.   The wavelength coverage of the band profile is set from 6.1 to 6.6 $\micron$ (following the same definition as described in \citet{Peeters02}). The fitted spectra of all sources composed of overlapping Gaussian profiles are shown in Figure \ref{subtracted}.

\subsection{Relationship between the 6.0 and 6.2 $\mu$m features}

\begin{figure}
\epsscale{1.0}
\plotone{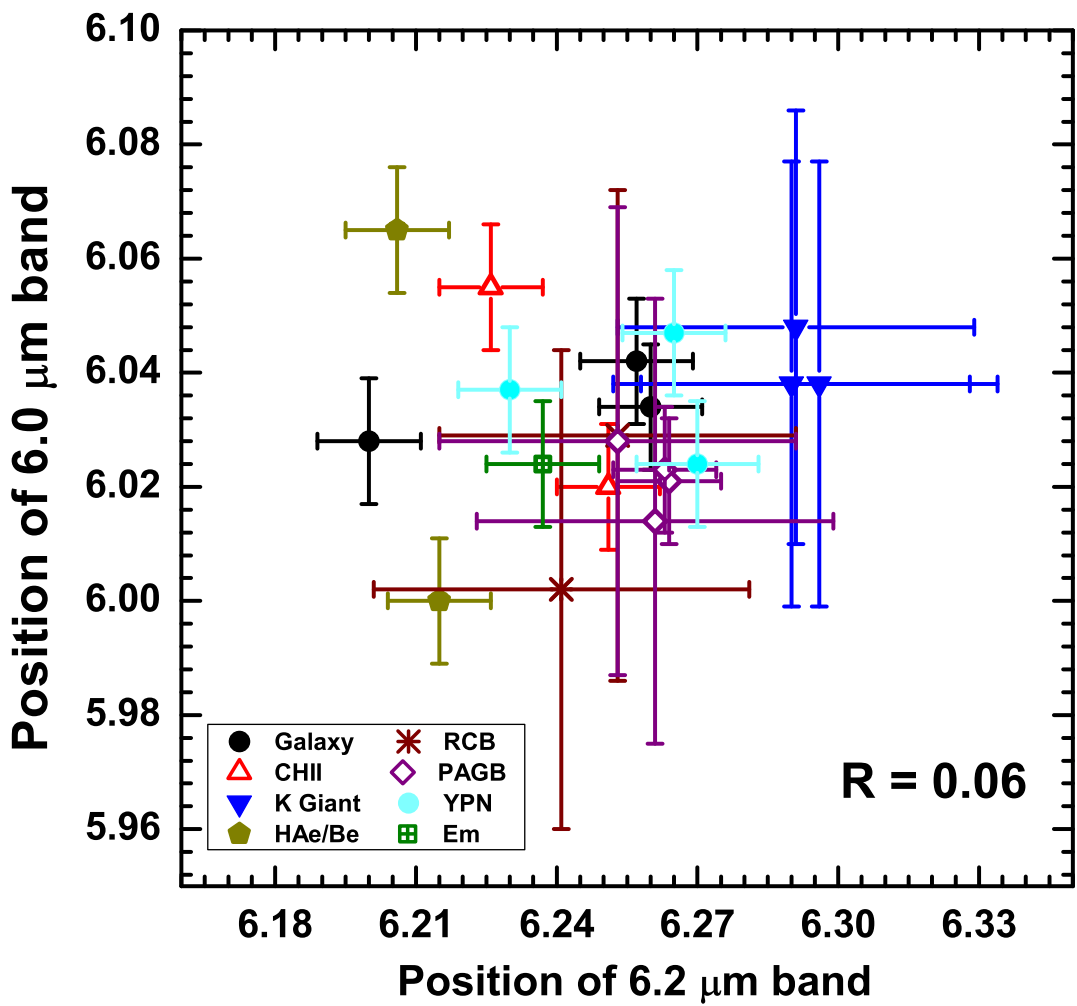}
\caption{The distribution of the central wavelength ($\lambda_{c}$) of the 6.2 $\micron$ and 6.0 $\micron$ bands showing the distributions of Galaxies (black 
filled circles), C\ion{H}{2} regions (red open triangles), K giants (blue filled triangles), HAe/Be stars (deep yellow pentagons), RCBs 
(brown asterisks), PAGBs (purple diamonds), YPNs (light blue filled circles), and Emission star (green squares). The correlation coefficient 
is given at the bottom right corner of the figure.}
\label{correlation}
\end{figure}

The 6.2 $\mu$m feature is widely observed in UIE sources but the 6.0 $\mu$m feature is less commonly present.  It would be interesting to see whether these features are related in any way.
If the 6.0 $\micron$ feature originates from similar stretching and bending mode as the 6.2 $\micron$ band\footnote{The 6.2 $\mu$m band has been commonly identified as due to aromatic C=C stretching mode, we find that it is actually a coupled feature of C=C stretching and C-H in-plane bending modes of aromatic compounds (Sadjadi et al. in preparation).}, 
there should exist a positive correlation between the positions of the 6.0 and 6.2 $\micron$ features as found for the positions of 6.2 and 7.7 $\mu$m features  \citep{Carpentier12}.  
In Figure ~\ref{correlation}, we explore the possible correlation between the central wavelength ($\lambda_{c}$) of the 6.0 $\micron$ and 6.2 $\micron$ bands. No clear correlation is found, suggesting  that the two features probably arise from different functional groups.
Figure \ref{correlation} shows that the peak wavelengths of 6.2 $\mu$m feature are clearly segregated among sources of different classes.  This may reflect the radiation environment in different classes of objects.  However, there is no such effect for the 6.0 $\mu$m feature.

\begin{figure}
\epsscale{1.0}
\plotone{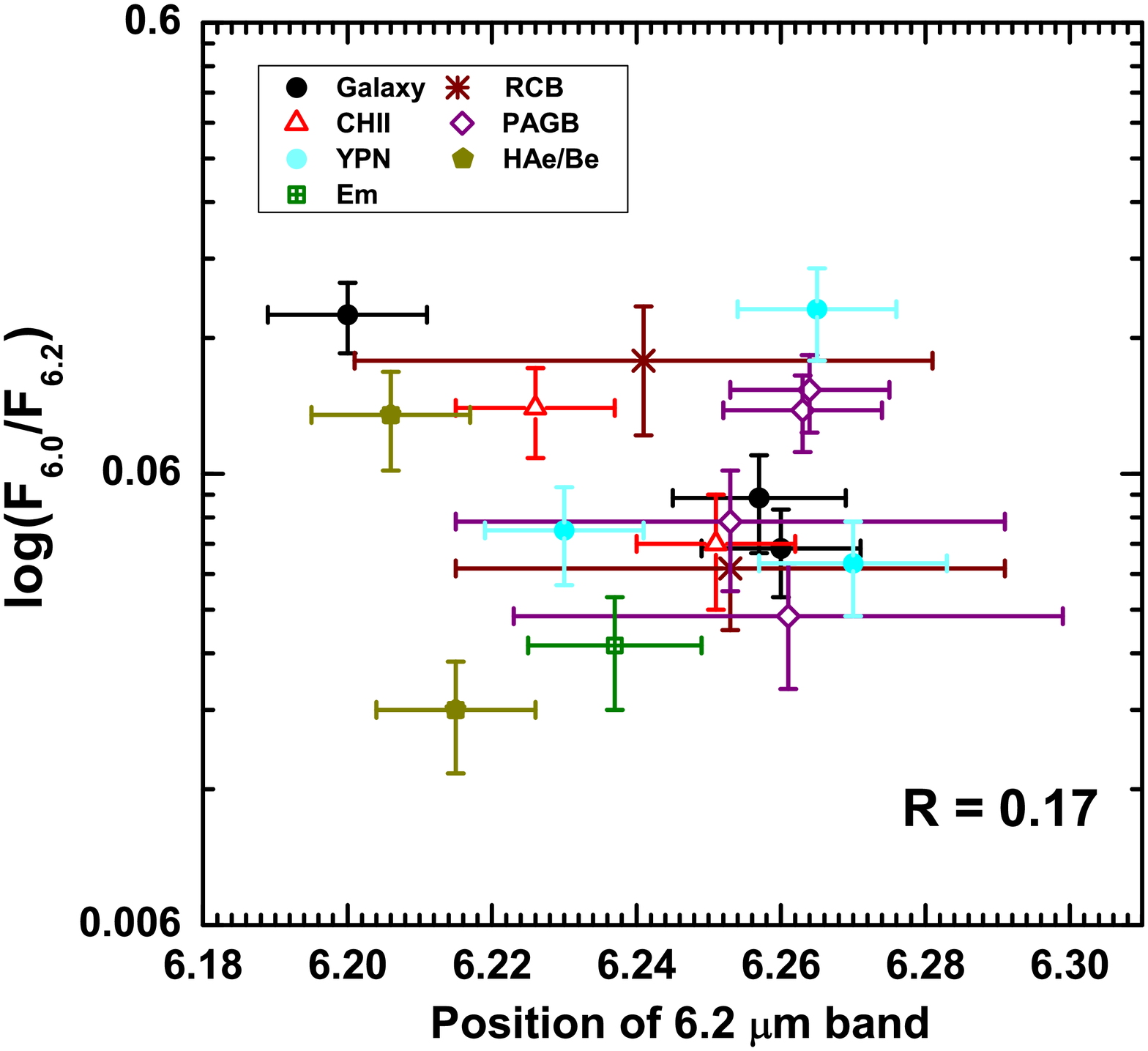}
\caption{The distribution of the 6.0 to 6.2 $\micron$ flux ratios. The fluxes of each feature are derived from the integrated strength of the
feature above the continuum. The symbols are the same in Fig.~\ref{correlation}. {The three blue points in Figure \ref{correlation} have been removed due to large flux uncertainties.}   The correlation coefficient is given at the bottom right corner of the figure.}
\label{fluxratio}
\end{figure}

Since the changing peak wavelengths of the 6.2 $\mu$m feature has been suggested to be the result of changing aliphatic to aromatic ratio  \citep{sloan2007, Pino08}, we also searched for possible correlation between the 6.0 to 6.2 $\mu$m flux ratios and the central wavelength of the 6.2 $\mu$m band.  No clear correlation is found (Figure \ref{fluxratio}).   

We also checked for possible correlations between 6.0 $\micron$ band central wavelength ($\lambda_{c}$) and the effective temperatures of the central sources (T$_{eff}$), flux ratio of 6.0 to 6.2 $\micron$ features, and the local radiation field ($G_{0}$)/electron density ($n_{e}$).   No obvious correlation is detected. The presence of 6.0 $\micron$ features does not seem to depend strongly on the  local physical conditions ($T_{eff}$, $G_{0}$, $n_{e}$) of the emitting regions . 

Previously, the  6.0 $\mu$m feature has been attributed to the C=O stretching mode \citep{st96,Peeters02}. However, we find that the 6.0 $\mu$m feature is detected in both C-rich and mixed chemistry objects, and C-rich objects do not show systematically stronger 6.0 $\mu$m/6.2 $\mu$m flux ratios than those having mixed chemistry (Table \ref{tab1}). 
This probably suggests that the carrier of the 6.0 $\mu$m feature is not strongly related to the abundance of oxygen, against the C=O hypothesis.

\section{Comparison with experimental data}

\begin{figure}
\epsscale{1.0}
\plotone{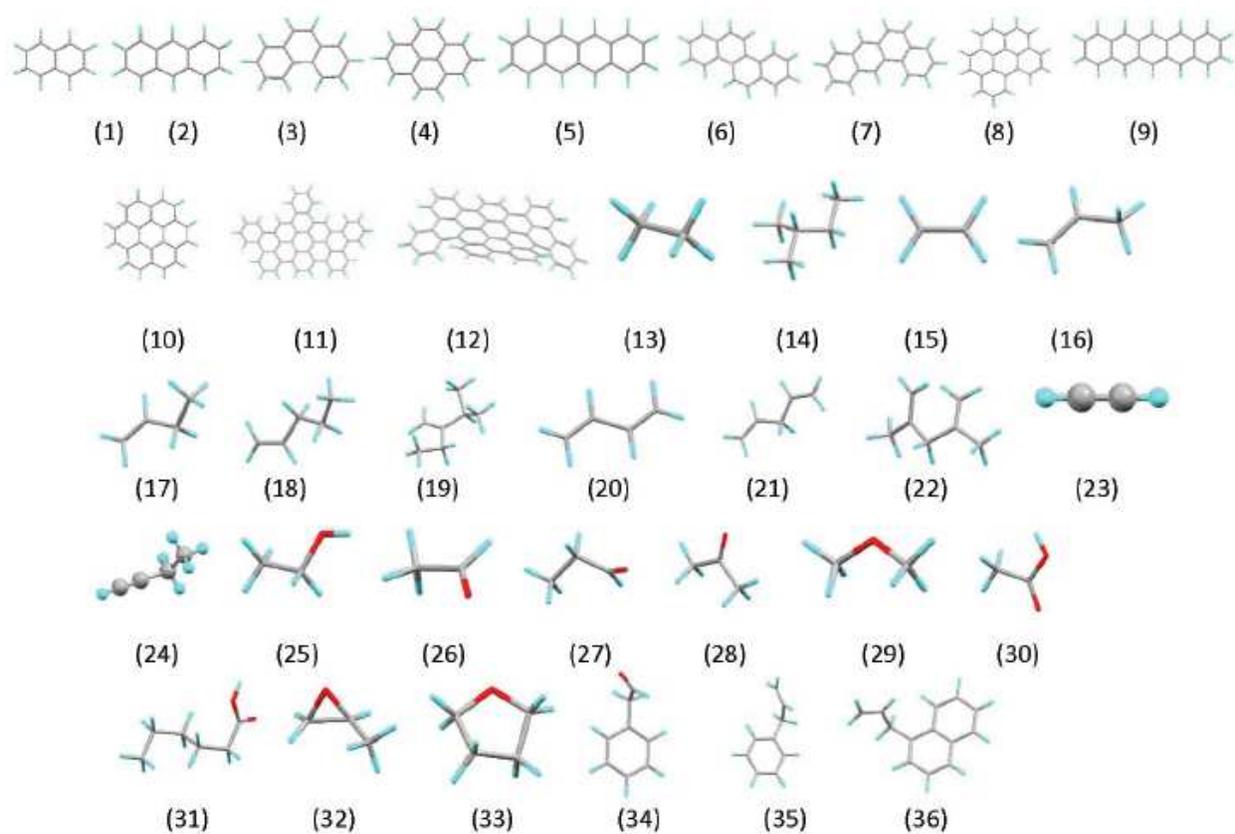}
\caption{Our sample of molecules searched within NASA and NIST experimental infrared spectra databases. Their corresponding names and chemical formula are listed in Table \ref{tab3}.  The C atoms are shown in grey, H atoms in blue, and O atoms in red.}
\label{sample}
\end{figure}

In order to determine the possible origin of the 6.0\,$\micron$ band, we investigate the theoretical and experimental spectra of a sample of molecules.  The names and chemical formulae of those molecules are listed in Table \ref{tab3} and their chemical structures shown in Figure \ref{sample} and \ref{theory_sample}. The NASA PAH database version 2.0 \citep{boersma2014} \footnote{http://www.astrochem.org/pahdb/} and NIST chemistry webbook (NIST Standard Reference Database Number 69) \footnote{http://webbook.nist.gov/chemistry/} are used as the main sources of experimental IR gas-phase data.

The 6.0 $\micron$ band has previously been suggested in the literature  to be due to the fundamental vibrational band of the carbonyl group (=C=O) found in ketone molecules   \citep{Peeters02}. Depending on the chemical structure of the compounds the stretching mode of C=O group appear in the range of 5.26 to 6.45 $\micron$ \citep[p.~390]{Colthup90}. However a closer look at the experimental data shows that stretching vibrations of C=N (5.92 to 6.13 $\micron$) and olefinic \footnote{According to the  IUPAC Gold Book "Olefins are acyclic and cyclic hydrocarbons having one or more carbon-carbon double bonds, apart from the formal ones in aromatic compounds. The class of olefins subsumes alkenes and cycloalkenes and the corresponding polyenes" \citep[p.~1353]{Moss95}.} C=C (5.95 to 6.25 $\micron$) groups also fall within the same wavelength range of the carbonyl group vibrations \citep[p.~390]{Colthup90}. In fact there are more olefinic C=C vibration modes clustered around  6 $\micron$ than =C=O and =C=N vibration modes.

\begin{figure}
\epsscale{1.0}
\plotone{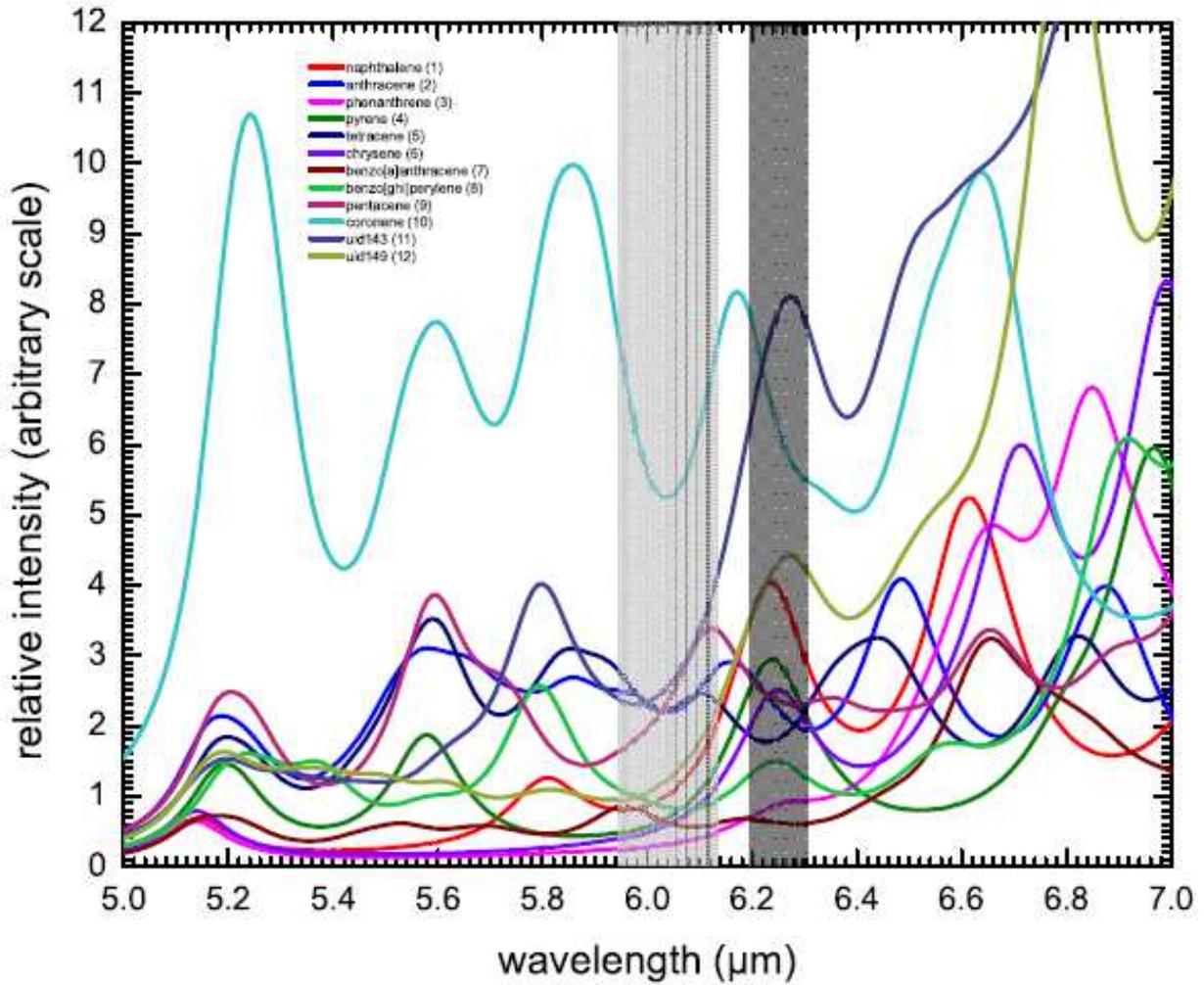}
\caption{Simulated 5--7 $\mu$m emission spectra derived from  experimental data of twelve PAH molecules (number 1--12 in Table \ref{tab3} and figure~\ref{sample}). Source: The NASA Ames PAH IR Spectroscopic Database.  The light gray bar marks the wavelength region 5.95--6.13 $\mu$m and the dark gray bar marks the 6.2--6.3 $\mu$m region.  The strong features around 5.2 and 5.8 $\mu$m have been attributed to overtone and combination bands of the aromatic C--H OOP bending modes at 11.2-11.3 $\mu$m.}
\label{pah_exp}
\end{figure}

Figure \ref{pah_exp} shows the simulated emission spectra assuming a thermal excitation (at 500\,K) model with a Drude profile from the experimental infrared absorption spectra of 12 neutral PAH molecules extracted from NASA data base.
Although these PAH molecules  show common features around 6.2 $\micron$, they do not exhibit the IR bands around 6.0 $\micron$. The only exception is a very weak band at 5.96 $\micron$ in benzo-a-anthracene spectra. 
Included in this plot are two large PAH molecules  C$_{48}$H$_{22}$ (UID143) and C$_{50}$H$_{22}$ (UID149). Their spectra also do not show a band at 6.0 $\micron$.  Some PAH molecules show strong features around 5.2 and 5.8 $\mu$m, which have been attributed to overtone and combination bands of the aromatic C--H out-of-plane (OOP) bending modes at 11.2-11.3 $\mu$m  \citep{alla1989, Boersma09}.  These spectra shown in Figure \ref{pah_exp} show that PAH molecules do not show consistent strong features in the 5 to 7 $\mu$m region, confirming previous calculations that the astronomical UIE spectra cannot be reproduced by combination of simple PAH molecules \citep{cook1998}. 

\begin{figure}
\epsscale{0.9}
\plotone{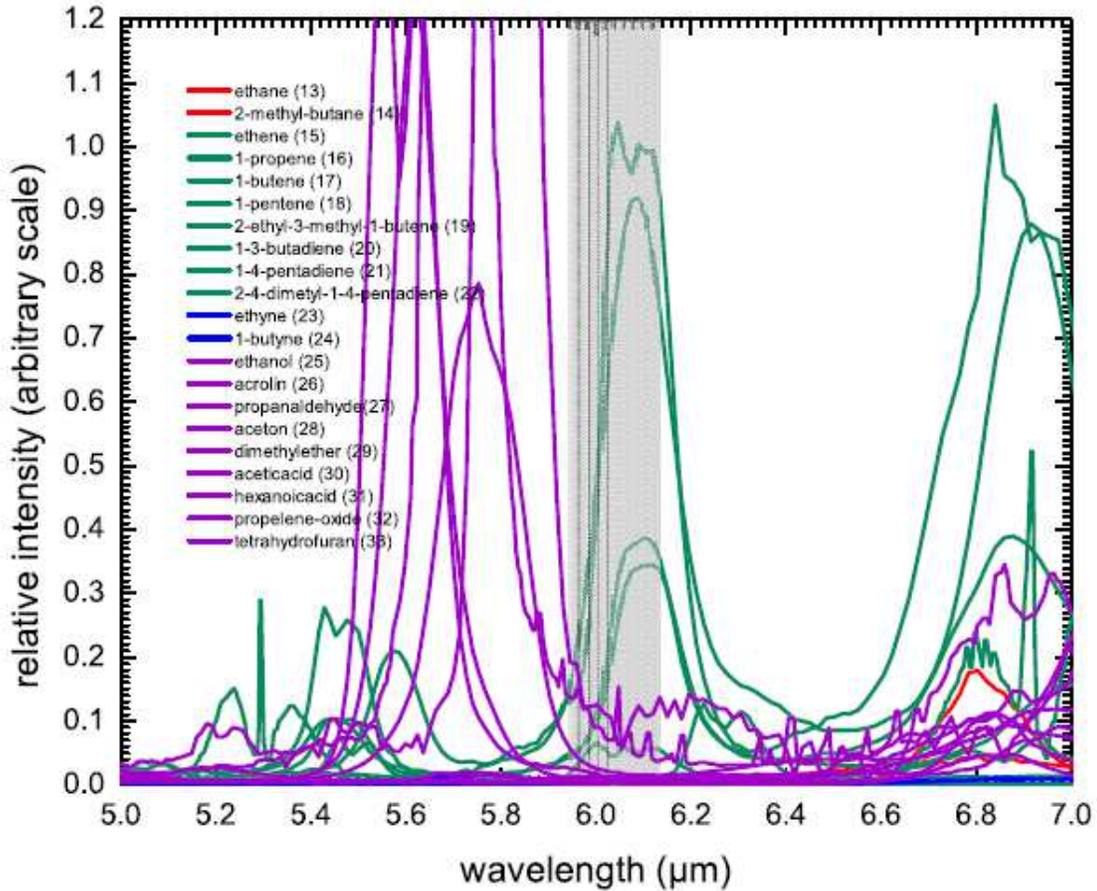}
\caption{Simulated 5--7 $\mu$m emission spectra derived from experimental data of twelve hydrocarbon molecules (number 13--24 in Table \ref{tab3} and figure~\ref{sample}) and nine O-containing hydrocarbon molecules with single, double and cyclic C--O bonds. (number 25--33 in Table \ref{tab3} and figure~\ref{sample}). Single C--C bond containing molecules in red, olefinic C--C double bond containing molecules in green, triple C--C bond in blue and all O-containing molecules in purple color. Except very weak feature around 5.8 $\mu$m for acetylenic species, these molecules do not show any absorption bands at this wavelength range of spectrum. Shaded area shows 5.95 to 6.13 $\mu$m wavelength range. Source: NIST Chemistry webbook (NIST Standard Reference Database Number 69).
 }
\label{olefin}
\end{figure}

Next we examine the gas-phase IR spectra of a number of simple olefins hydrocarbons with one or two isolated or conjugated double bonds (Figure \ref{olefin}). Also shown in this plot are the IR spectra of simple alkanes (CC single bond) and alkynes (CC triple bond).   Figure \ref{olefin} shows that among the hydrocarbons with different kinds of CC bonds, only olefins (green lines) show  features within the range of 6 $\micron$ to 6.2 $\micron$.   For comparison with the spectra of olefins, we have plotted the experimental gas-phase IR spectra of a number of simple O-containing hydrocarbons (purple lines in Figure \ref{olefin}). These molecules include  alcohols (C$-$O$-$H), ketones (C=O), ethers (C$-$O$-$C), carboxylic acids ($-$COOH), epoxides and furans (cyclic C$-$O$-$C). None of these molecules  (with different C--O bonds) show the feature around 6.0 $\micron$.   Instead they show strong features around 5.5, 5.6, 5.8 and 5.9 $\micron$, with the carboxylic acids bands at the most blue-shifted part. Propylene oxide and tetrahydrofuran are almost inactive. 

\section{Quantum chemistry calculations}

In this section, we explore the effects of olefins and carbonyls  sidegroups attached to a variety of aromatic molecules. We use BHandHLYP hybrid functionals  \citep{Becke93} in combination with polarization consistent basis set PC1  \citep{Jensen01, Jensen02} to obtain the equilibrium geometries and the fundamental vibrational harmonic frequencies of the molecules. The calculations are based on density functional theory (DFT) using the Gaussian 09, Revision C.01 software package  \citep{Frisch09} running on the HKU grid-point supercomputer facility. The double-scaling-factors scheme  \citep{Laury12} are then applied to the DFT vibrational frequencies. Specifically, the vibrational frequencies $>$1000 cm $^{-1}$ and $<$1000 cm$^{-1}$  are scaled by 0.9311 and 0.9352 for BHandHLYP hybrid functionals. The astronomical infrared emission spectra are then simulated by assuming a thermal excitation (at 500K) model with a Drude profile. The good accuracy and reliability of such DFT/Drude modeling in simulation of astronomical infrared emission bands was demonstrated in our previous study  \citep{Sadjadi15a}. 

All geometries have been optimized under the default criteria of cited ab intio quantum chemistry package. The optimized geometries are all characterized as local minima, established by the positive values of all frequencies and their associated eigenvalues of the second derivative matrix. Visualizing and manipulating the results of vibrational normal mode analysis were performed by utilizing the Chemcraft program.

\begin{figure}
\epsscale{1.0}
\plotone{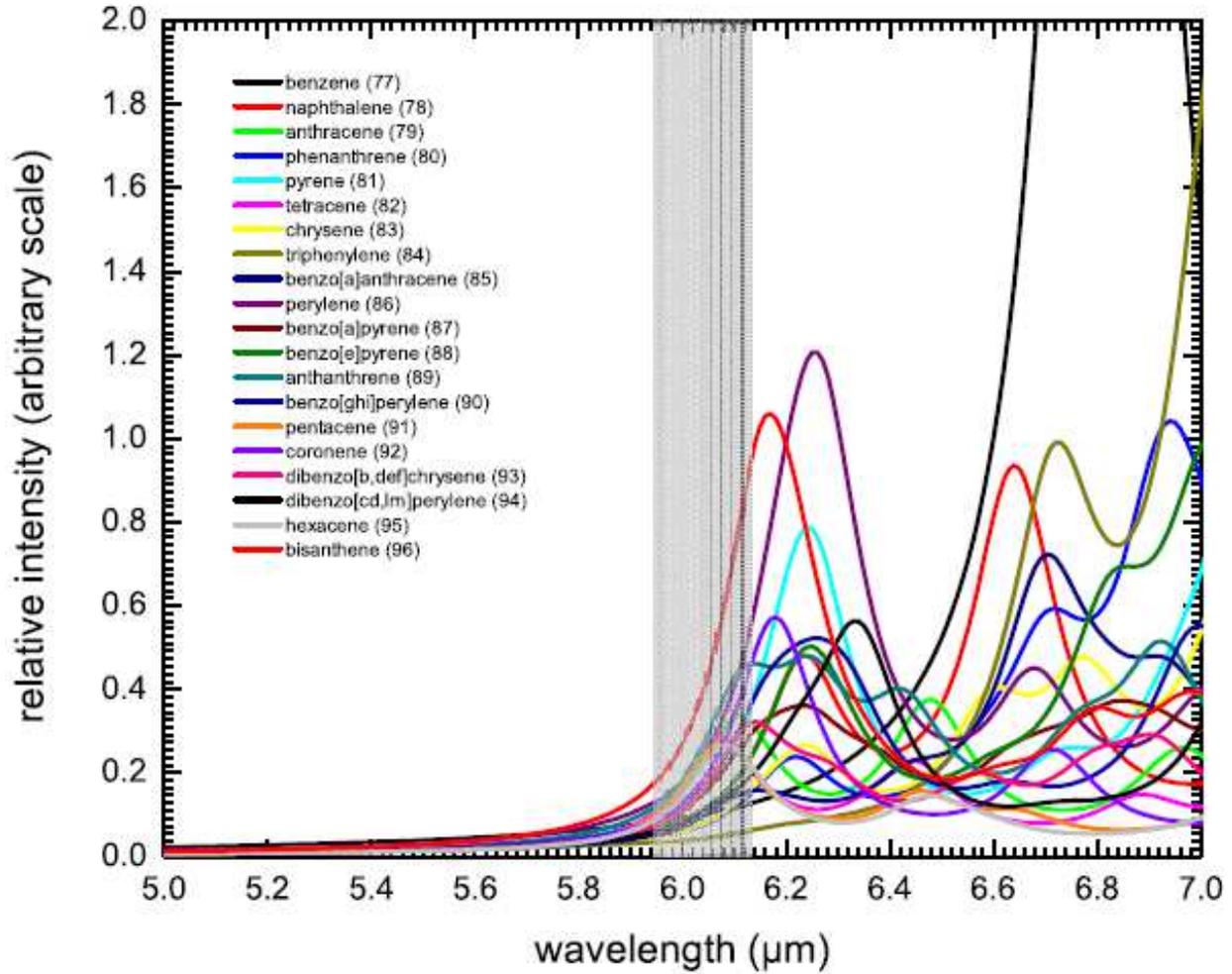}
\caption{Theoretical simulated 5--7 $\mu$m emission spectra  for twenty classical PAH molecules (number 77--96 in Table \ref{tab3} and Figure \ref{theory_sample}). These molecules represent the  core of the allyl and aldehyde structures in Table \ref{tab3} and Figure \ref{theory_sample}. 
}
\label{pah_theory}
\end{figure}

\begin{figure}
\epsscale{0.8}
\plotone{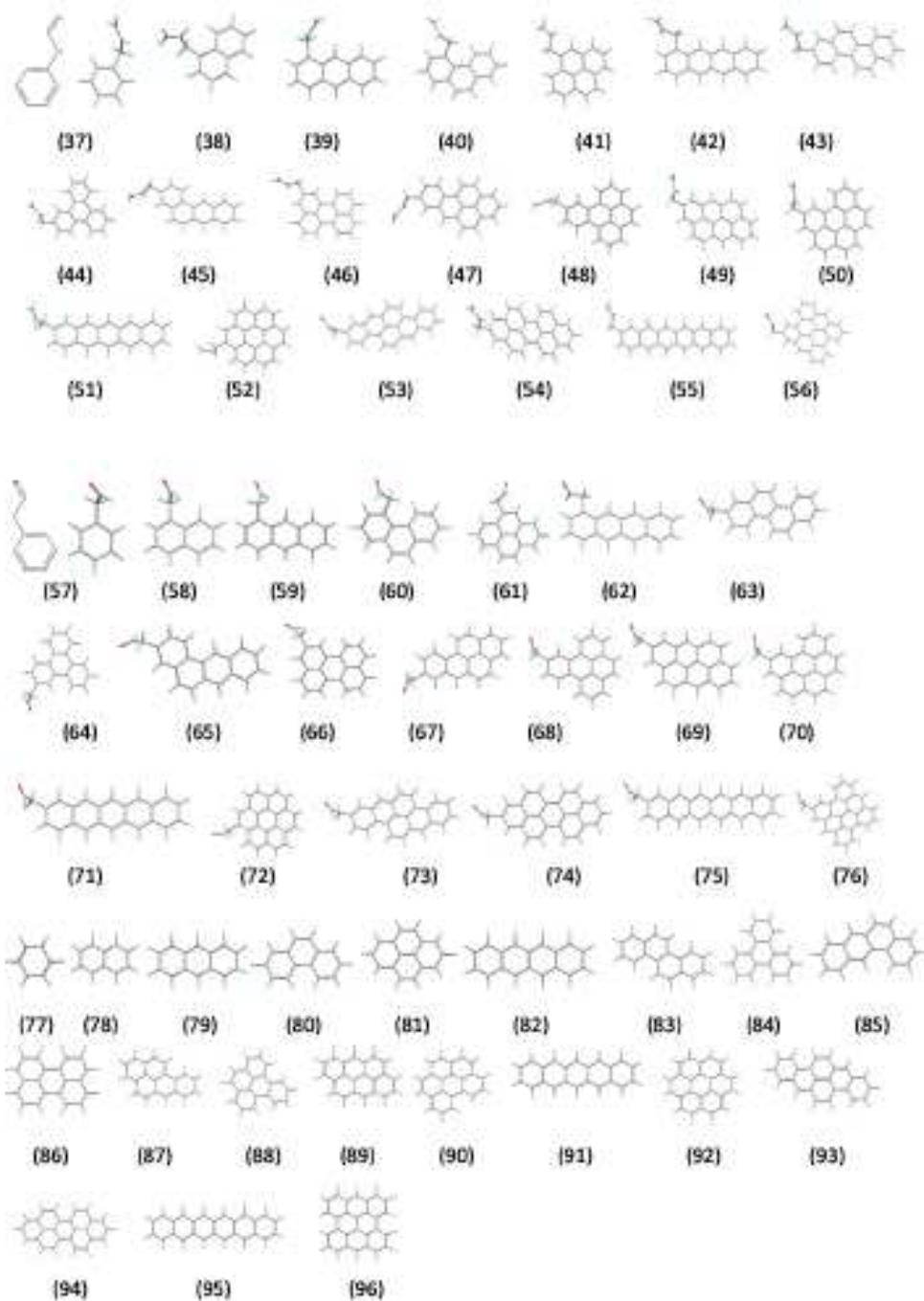}
\caption{The local minimum geometries of allyl-aromatic (number 37--56 in Table \ref{tab3}, Group A) and aldehyde-aromatic (number 57--76 in Table \ref{tab3}, Group B) molecules calculated with the BHandHLYP/PC1 quantum chemical model.  The C atoms are shown in grey, H atoms in blue, and O atoms in red}
\label{theory_sample}
\end{figure}

\begin{figure}
\epsscale{1.0}
\plottwo{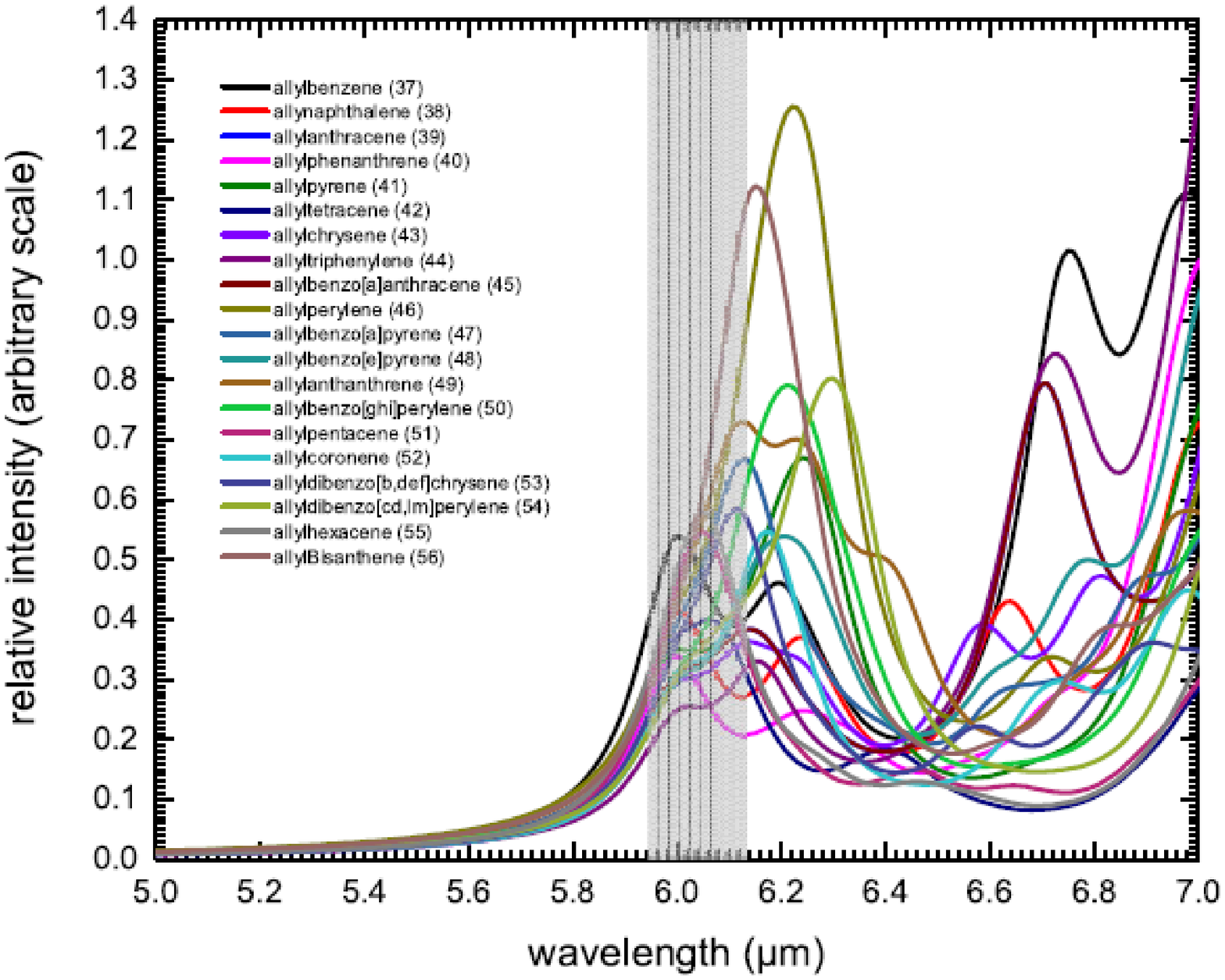}{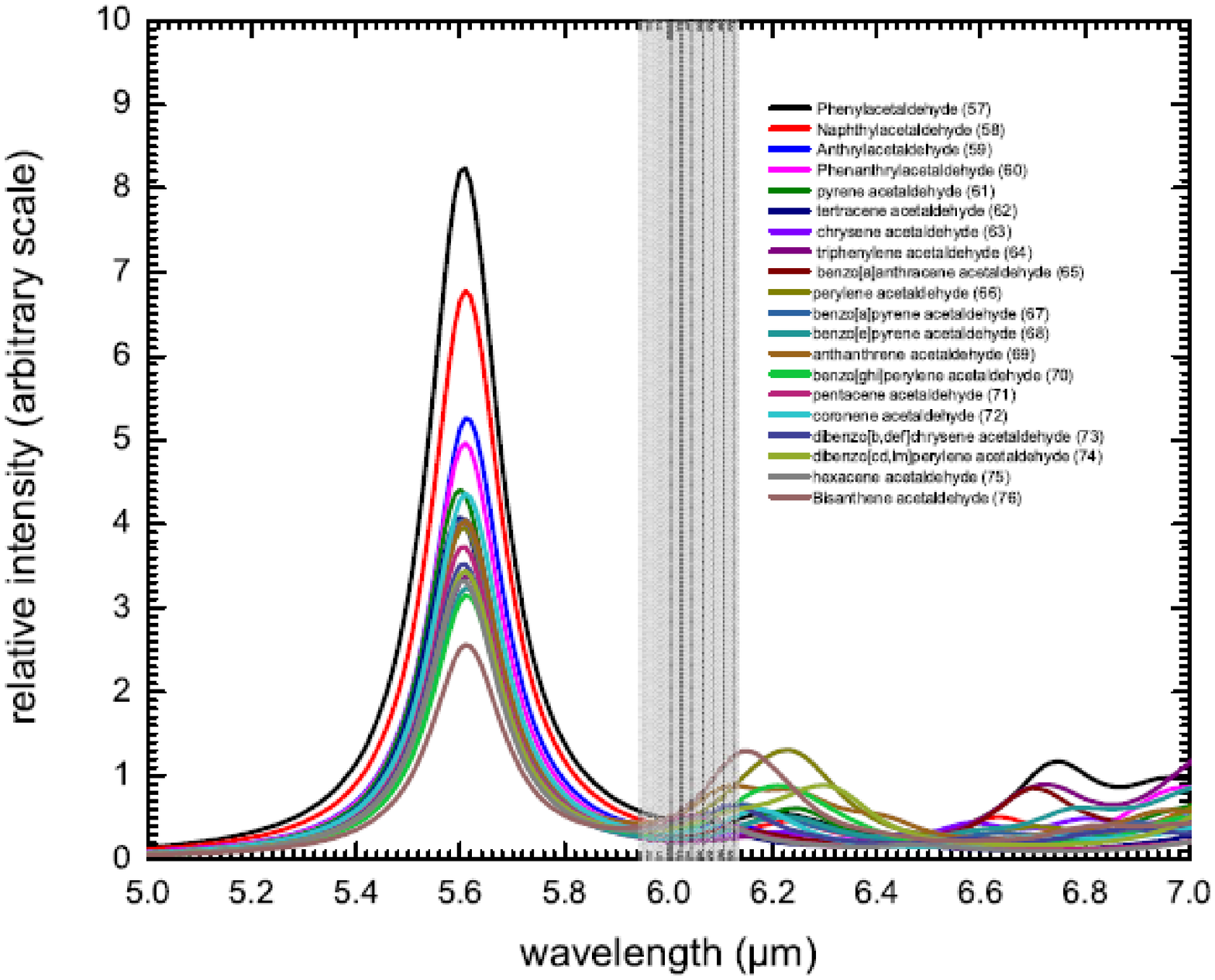}
\caption{Theoretically simulated 5--7 $\mu$m emission spectra for twenty allyl-aromatic molecules, group A (number 37--56 in Table \ref{tab3} and Figure \ref{theory_sample}) (left panel) and twenty aldehyde-aromatic molecules, group B (number 57--76 in Table \ref{tab3} and Figure \ref{theory_sample}) (right panel).  The shaded area shows the 5.95--6.13 $\mu$m region.
}
\label{groupAB}
\end{figure}

We begin by using 20 of the 60 PAH molecules studied by  \citet{Sadjadi15a}  as aromatic cores and study the effect of adding peripheral side groups to these cores. The theoretical spectra of these 20 PAH molecules are shown in Figure \ref{pah_theory}. No 6.0 $\mu$m feature is found in any of the spectra of these 20 PAH molecules. Olefins and carbonyl functional groups are added to these molecules to create two classes of mixed aromatic- aliphatic molecules. Group A consists of PAH molecule with $-$CH$_2$HC$=$CH$_2$ functional groups (number 37--56 in Table \ref{tab3} and Figure \ref{theory_sample}, and Group B consists of PAH molecules with $-$CH$_{2}$HC=O functional group (number 57--76 in Table \ref{tab3} and Figure \ref{theory_sample}). The simulated spectra of group A and B are presented in Figure \ref{groupAB}.   Figure  \ref{groupAB} shows that the addition of olefin side group to PAH molecules (group A) can potentially produce a feature at 6.0 $\mu$m along with a 6.2 $\mu$m feature. However, no feature at 6.0 $\mu$m is observed in the spectra of carbonyl-bearing compounds (group B).

We note that number of aromatic rings among the Group A molecules ranges from 1 to 8 but the number of  olefinic functional groups remains fixed at one.  This is reflected in the changing 6.2 to 6.0 $\mu$m feature among the group A molecules.

\begin{figure}
\epsscale{1.0}
\plotone{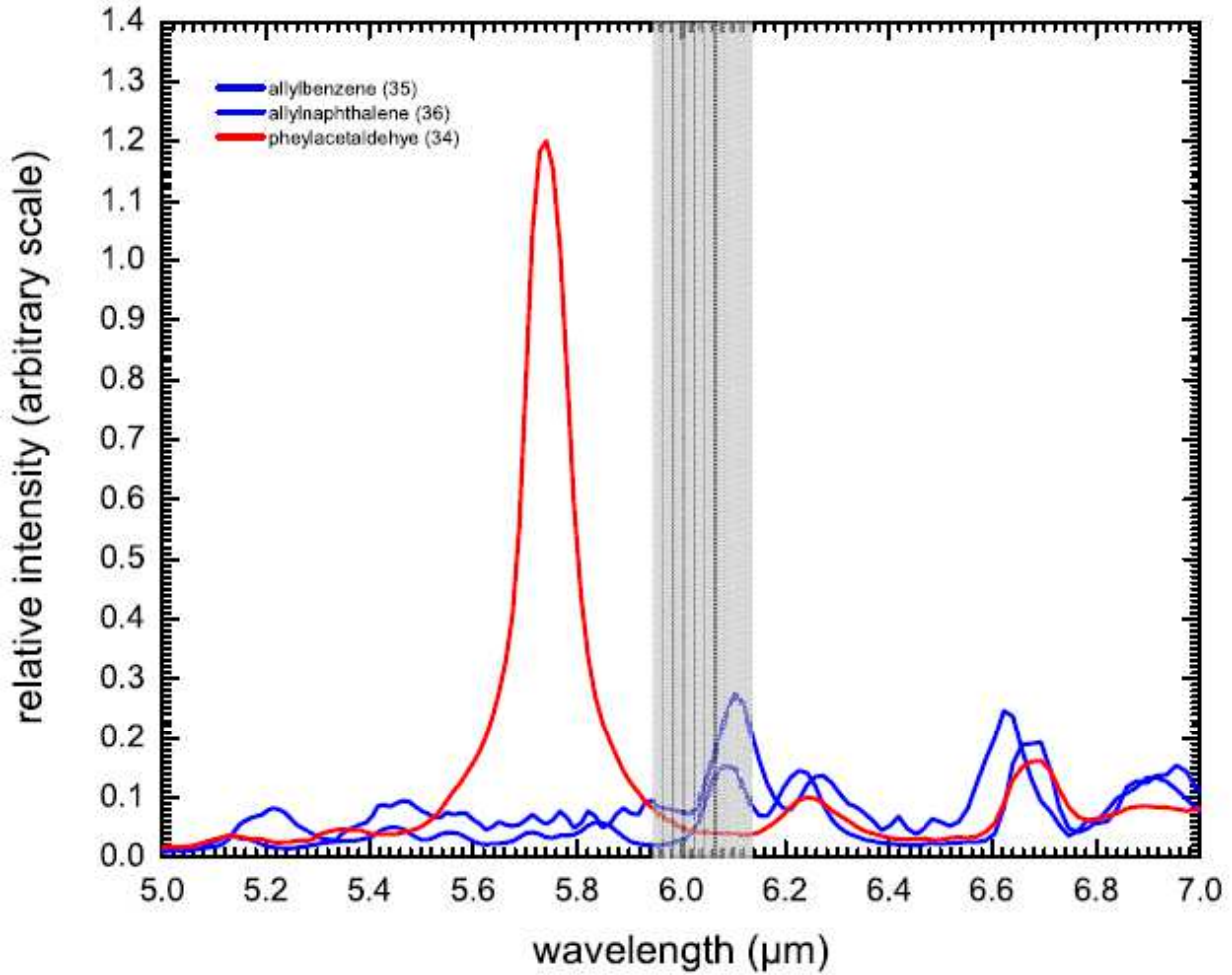}
\caption{Simulated 5--7 $\mu$m emission spectra from experimental data of three molecules studied by theory. The blue lines correspond to  allyl compounds (number 35 and 36 in Table \ref{tab3} and Figure \ref{theory_sample}) and the red line is the aldehyde (number 34 in Table \ref{tab3} and Figure \ref{theory_sample}).  The shaded area shows the 5.95--6.13 $\mu$m region.   Source: NIST Chemistry webbook (NIST Standard Reference Database Number 69).
}
\label{exp_ole}
\end{figure}

As a check on our theoretical calculations, we  plotted in Figure \ref{exp_ole} the experimental gas-phase IR spectra of two olefin (Group A) and one carbonyl (group B) molecules. We can see that the two mixed aromatic-olefin compounds have features within the range of 6.0--6.2 $\micron$ whereas the mixed aromatic-carbonyl compound does not show such a feature.  Instead it shows a sharp band at 5.7 $\micron$. These results are consistent with our theoretical results.

\begin{figure}
\epsscale{1.0}
\plotone{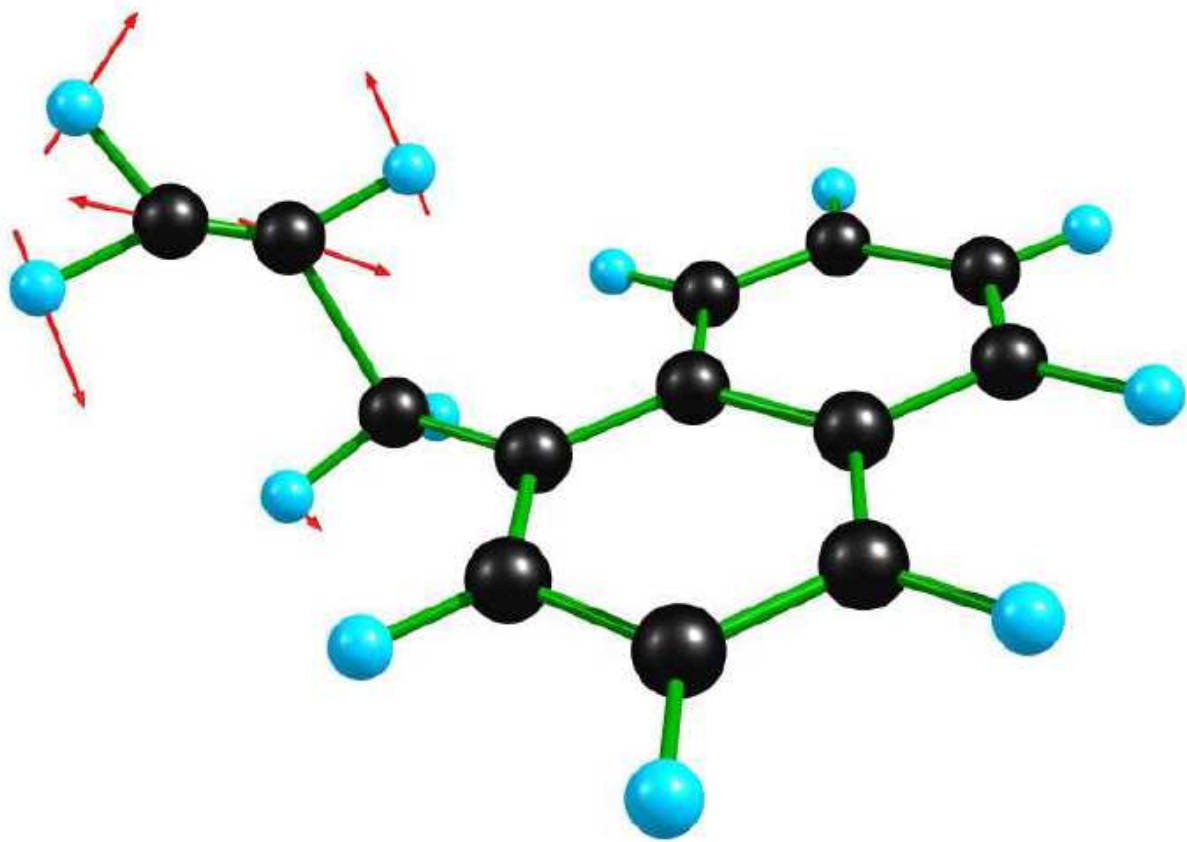}
\caption{The displacement vectors of the normal mode vibrational motion at 6.0 $\micron$ in allynaphthalene (C$_{13}$H$_{12}$, number 38 in Table \ref{tab3} and Figure \ref{theory_sample}).  An animation can be viewed in the on-line version of the Journal.  Both C=C and C$-$H bending modes are present.}
\label{vibrations}
\end{figure}

Table \ref{tab4} summarizes the central wavelength ($\lambda_{c}$) of the  6.0 $\micron$ and 6.2 $\micron$ features in the theoretical spectra (Figure \ref{groupAB}) of the group A molecules. 
All group A molecules show transitions around 6.0 $\mu$m  but after the application of a broadening profile, some of these features are only seen as part of the shoulder of the stronger 6.2 $\mu$m feature.  
A distinct peak around 6.0 $\micron$  is seen in about half of the Group A molecules. The vibrational analysis of \citet{Sadjadi15b} shows that the 6.0 $\mu$m feature is produced by olefinic C=C stretching coupled with aliphatic C--H bending modes (Figure \ref{vibrations}). It is therefore possible that  aliphatic fragments with olefinic groups can explain the observed astronomical UIE band at 6.0 $\micron$.

\section{Formation of olefinic groups in hydrogenated aromatic compounds}

\begin{figure}
	\begin{center}
	\includegraphics[width=30pc]{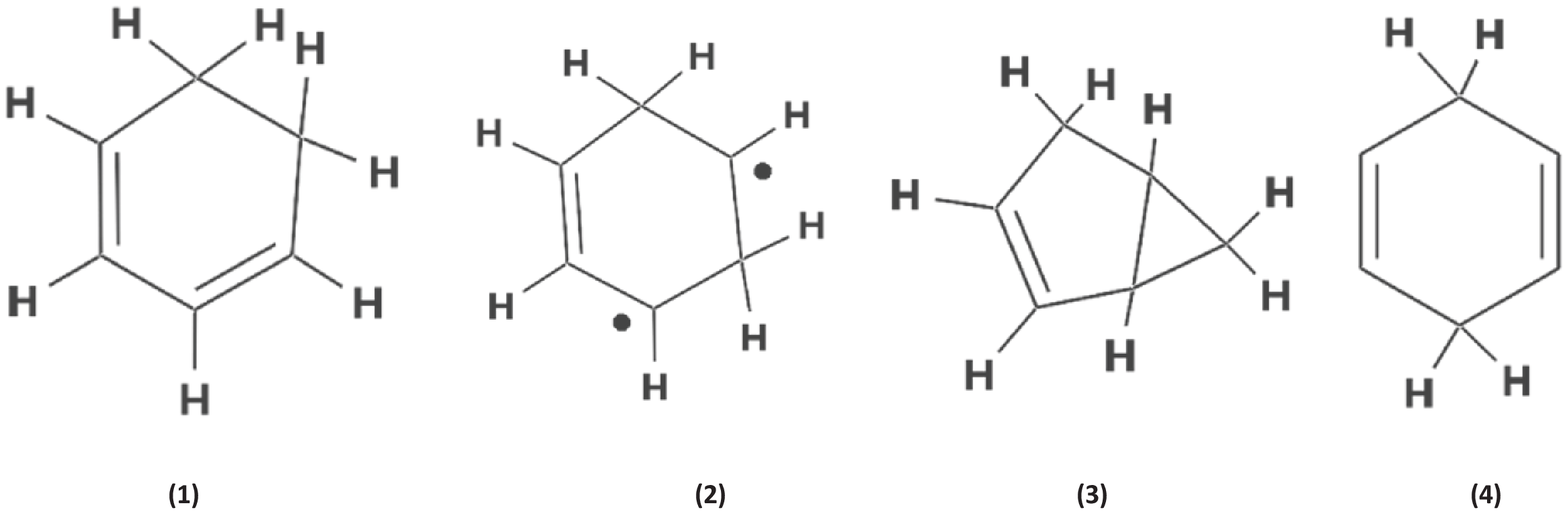}
	\end{center}
	\caption{\label{mixed1}The 2D resonance model of hydrogentated benzenes.}
\end{figure}

Hydrogenation of molecules with conjugated $\pi$-systems such as benzene, PAHs and fullerenes, could result in the formation of olefinic C=C in their molecular skeleton. This can be shown in  resonance model 2D diagrams for benzene molecules  (Figure \ref{mixed1}). If the hydrogenation occurs at meta and para positions, this would lead to the formation of olefinic C=C bonds (Figure \ref{mixed1}: structures 2, 3 and 4).   If hydrogenation occurs at ortho position the conjugation of $\pi$-system in benzene ring is partially preserved (Figure \ref{mixed1}: structure 1). Note that structure 2 is a biradical open shell molecule with two unpaired electrons. We assume that this species has a very short life time and convert to structure 3 with closed shell electronic structure.

\begin{figure}
	\begin{center}
	\includegraphics[width=30pc]{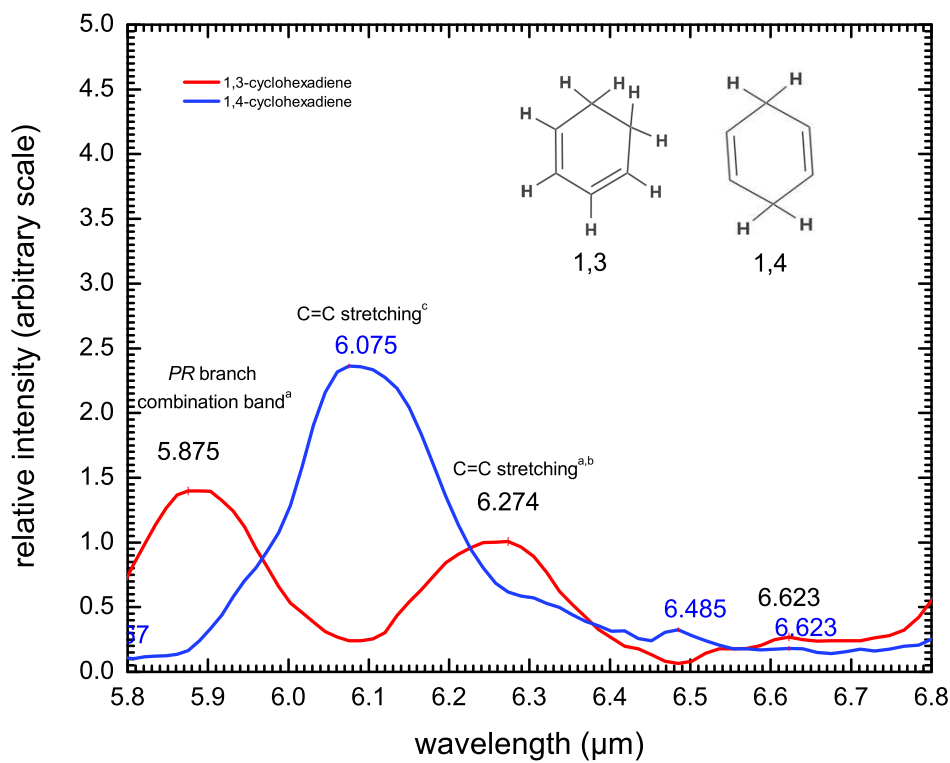}
	\end{center}
	\caption{\label{mixed2}Simulated 5--7 $\mu$m emission spectra from experimental data for two cyclohexadiene isomers. The blue curve is 1,4 and the red curve is 1,3 cyclohexadiene ($^a$ \citet{Lauro1969}, $^b$ \citet{Autrey2001}, $^c$ \citet{Moon1998}).}
\end{figure}

The experimental gas-phase IR spectra of structures 1 (1,3 cyclohexadiene) with two conjugated C=C bonds shows two bands at 6.274 $\mu$m and 5.875 $\mu$m (Figure \ref{mixed2}). They are assigned to C=C stretching  and  combination band respectively  \citep{Lauro1969,Autrey2001}. Structure 4 (1,4 cyclohexadiene) with two non-conjugated C=C bonds has an IR band at 6.075 $\mu$m as shown in Figure \ref{mixed2}. This band is also assigned to C=C stretching by applying both DFT and ab initio correlated methods \citep{Moon1998}.
Thus the 6 $\mu$m features not only signifies the presence of olefinic C=C bonds but also shows the position of the double bonds within the skeleton. Such information can be applied to the identification of molecular structure in hydrogenated aromatic compounds. 

\begin{figure}
	\begin{center}
		\includegraphics[width=30pc]{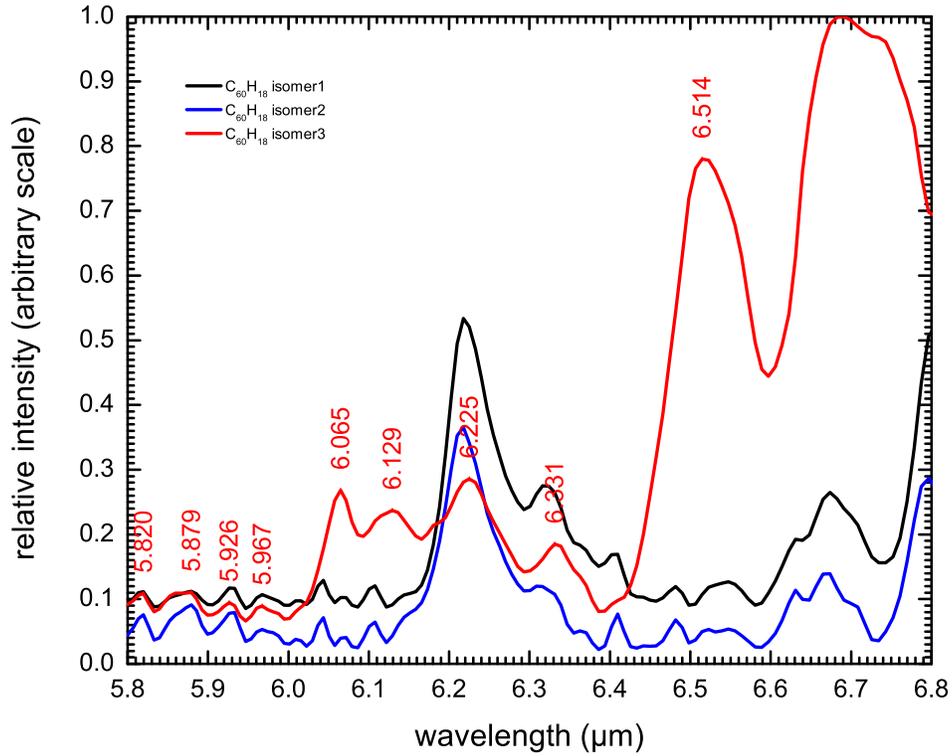}
	\end{center}
	\caption{\label{mixed3}Simulated 5--7 $\mu$m emission spectra from experimental data for three C$_{60}$H$_{18}$ isomers at 50$^\circ$ C in solid phase. Original Data from \citet{Cataldo2012} and Cataldo, private communications.} 
	\end{figure}

Another example to search for similar IR signatures are hydrogenated fullerenes (fulleranes).
In their series of experimental studies on fulleranes, Cataldo's group has synthesized different isomers of C$_{60}$H$_{18}$ and reported their FTIR spectra  \citep{Cataldo2012}. The FTIR spectra of three of  these isomers in solid state are presented in Figure \ref{mixed3}.
Figure \ref{mixed3} shows that isomer 3 has two extra bands at 6.065 and 6.129 $\mu$m in addition to the other bands at 6.225 and 6.331 $\mu$m. We suggest that the appearance of the band at 6.065 $\mu$m can be attributed to the formation of olefinic C=C in the structure of this isomer during the hydrogenation. Since only the hydrogenation of certain carbon atoms in fullerene cage can lead to olefinic C=C, as we have shown for the case of benzene,  the choice for possible structures for isomers will be restricted. This will significantly help the identification of the correct molecular structure of this isomer. This discussion is also applicable to hydrogenated PAH molecules.

\section{Conclusions}

In this paper, we present experimental and theoretical spectra of olefins for comparison with astronomical spectra showing the 6.0 $\mu$m UIE feature.  Comparisons are also made with pure aromatic (PAH) molecules and molecules containing the carbonyl group.  The advantages of the assumption of the 
olefinic hypothesis can be summarized as follows:
 
\begin{itemize}
  \item {The wavelength range of the C=C vibrational mode (5.96--6.25 $\micron$) measured experimentally is closer to the astronomically observed wavelength peak of the 6.0 $\micron$ feature than the C=O vibrational mode.} 

\item{Theoretical calculations show that the 6.0 $\micron$ feature can generally arise from simple hydrocarbon molecules with olefinic double-bond functional groups. }

\item{The olefinic C=C are inherently not a strong band, so there is no need to invoke the extra assumption of low concentration to explain the weak intensity of 6.0 $\micron$ features. This assumption can better explain why such a feature is common in different objects with different chemistry. }


\end{itemize}

We note that the coming {\it James Webb Space Telescope (JWST)} will offer higher sensitivity and spatial resolution which can be used to increase the sample size and better identify the nature of the 6.0 $\mu$m band. 

The UIE phenomenon is a complex one, consisting of major UIE bands and minor features as well as broad emission plateaus.  It is extremely unlikely that this phenomenon can be explained by simple gas-phase, purely aromatic PAH molecules.   This study is part of an exercise to learn more about the nature of vibrational bands of carbonaceous compounds with mixed aromatic/aliphatic structures.  We show that the presence of olefinic side groups can lead to minor features in UIE sources.  Further work is needed to fully explore the effects on aliphatic side groups as possible contributors to the UIE phenomenon.

{\flushleft \bf Acknowledgements~}
We thank an anonymous referee for his/her many helpful comments which led to improvements in this paper.  
The Laboratory for Space Research was established by a special grant from the University Development Fund of the University of Hong Kong.  This work is also in part supported by grants from the HKRGC (HKU 7027/11P and HKU7062/13P.).

\clearpage


\begin{figure*}
\begin{center}
\begin{tabular}{c}
\resizebox{140mm}{!}{\includegraphics{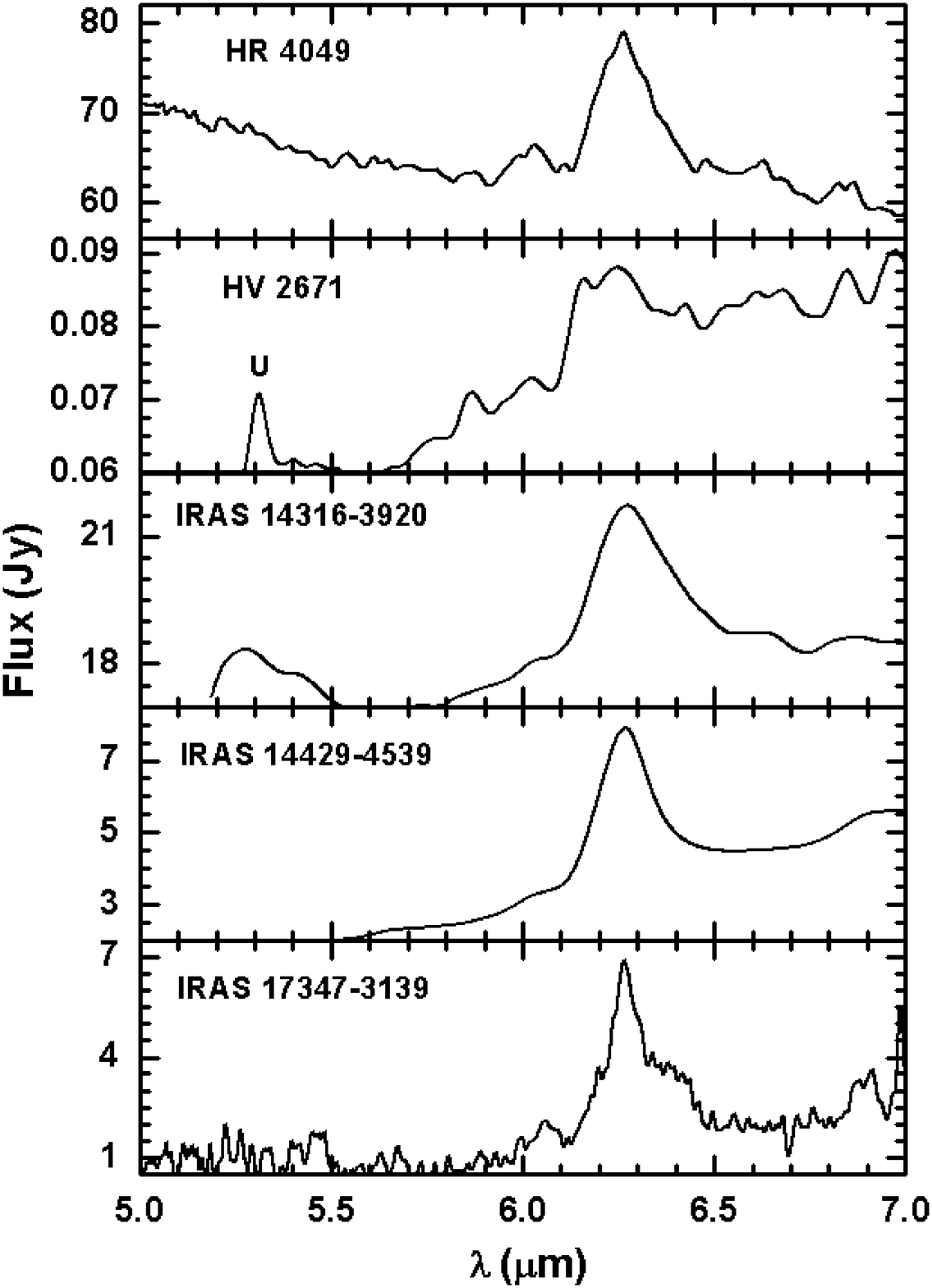}} \\
\end{tabular}
\end{center}
\end{figure*}

\begin{figure*}
\begin{center}
\begin{tabular}{c}
\resizebox{140mm}{!}{\includegraphics{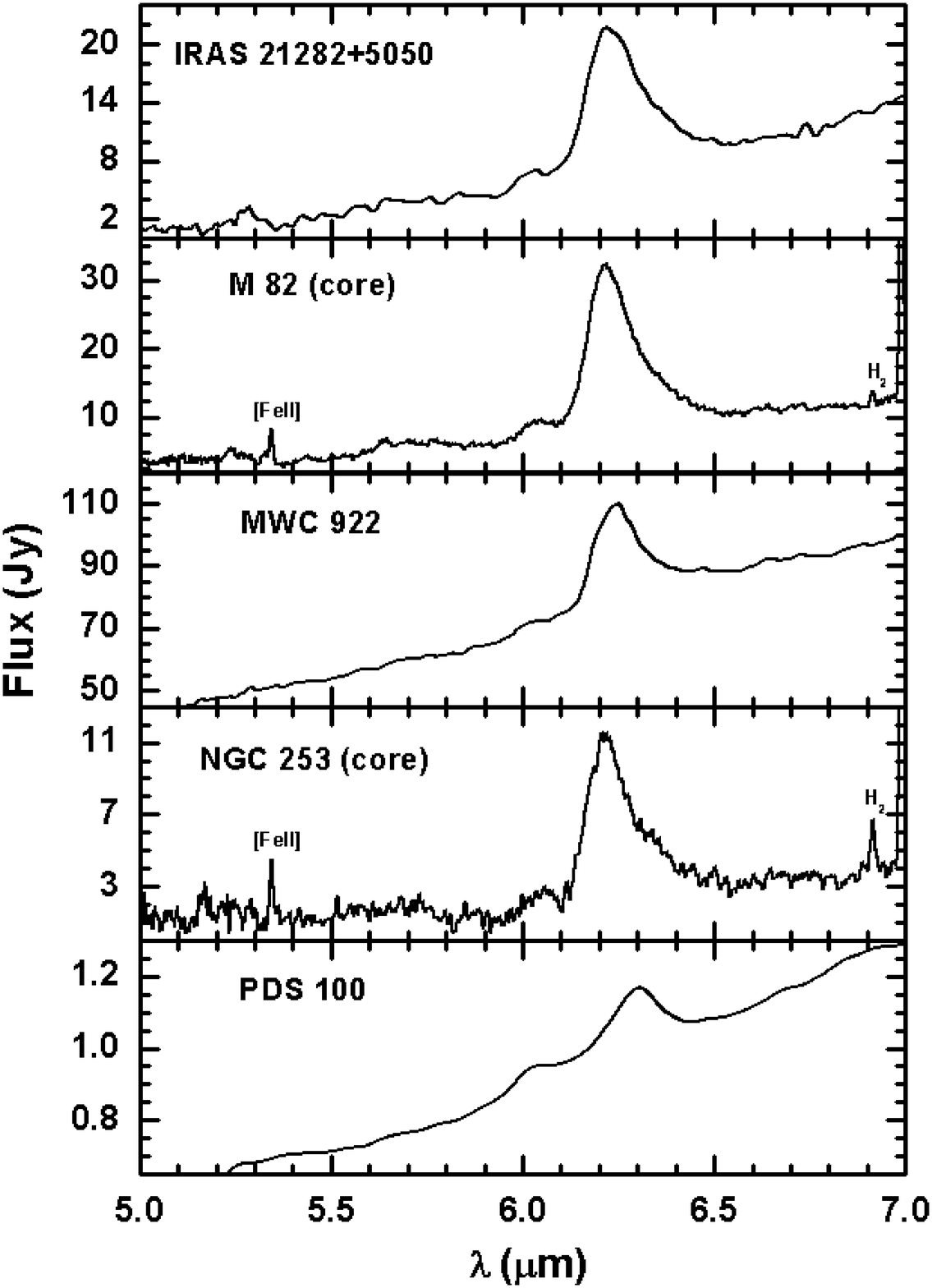}} \\
\end{tabular}
\end{center}
\end{figure*}

\begin{figure*}
\begin{center}
\begin{tabular}{c}
\resizebox{140mm}{!}{\includegraphics{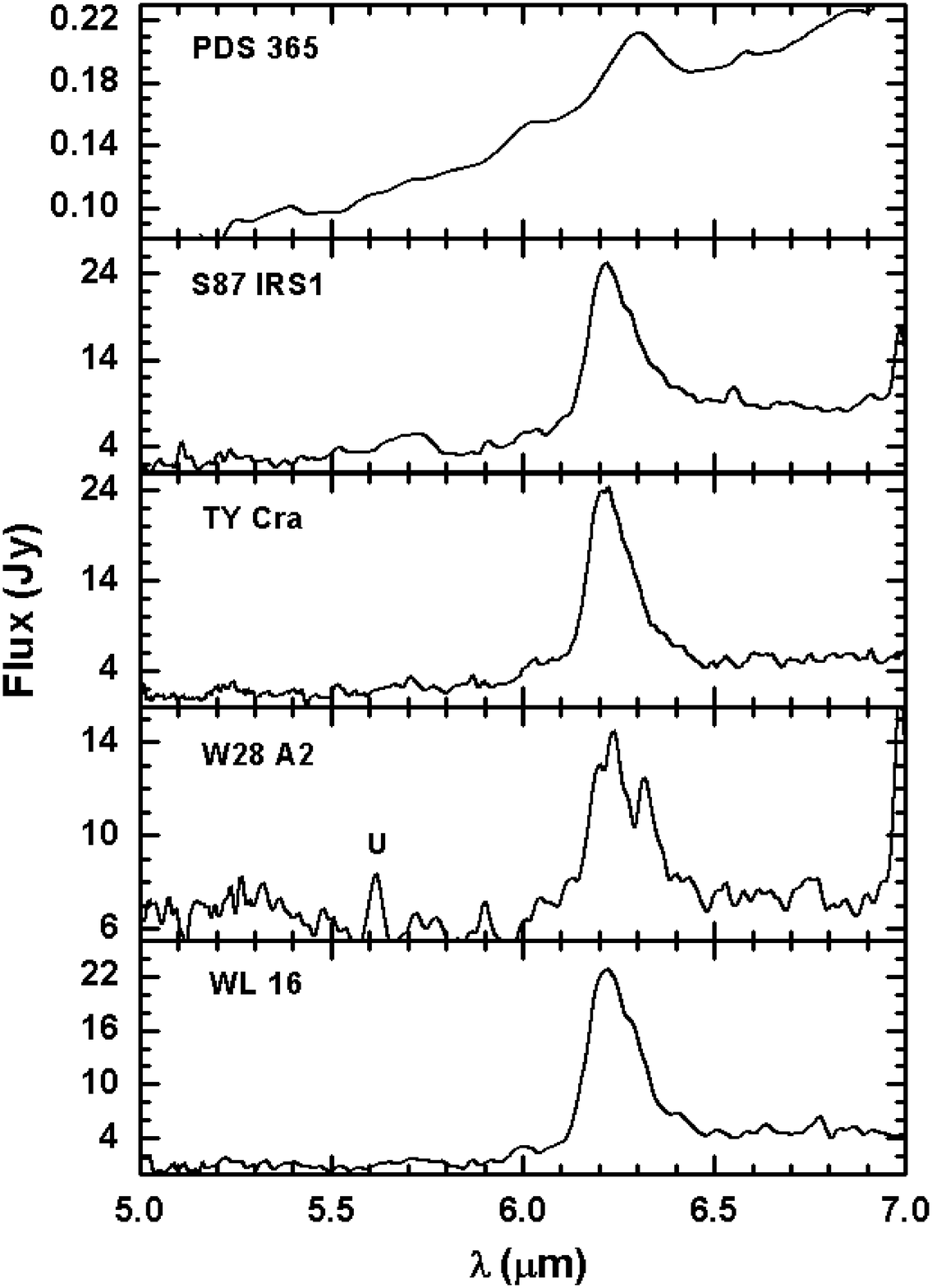}} \\
\end{tabular}
\end{center}
\end{figure*}

\clearpage

\begin{figure*}
\begin{center}
\begin{tabular}{c}
\resizebox{140mm}{!}{\includegraphics{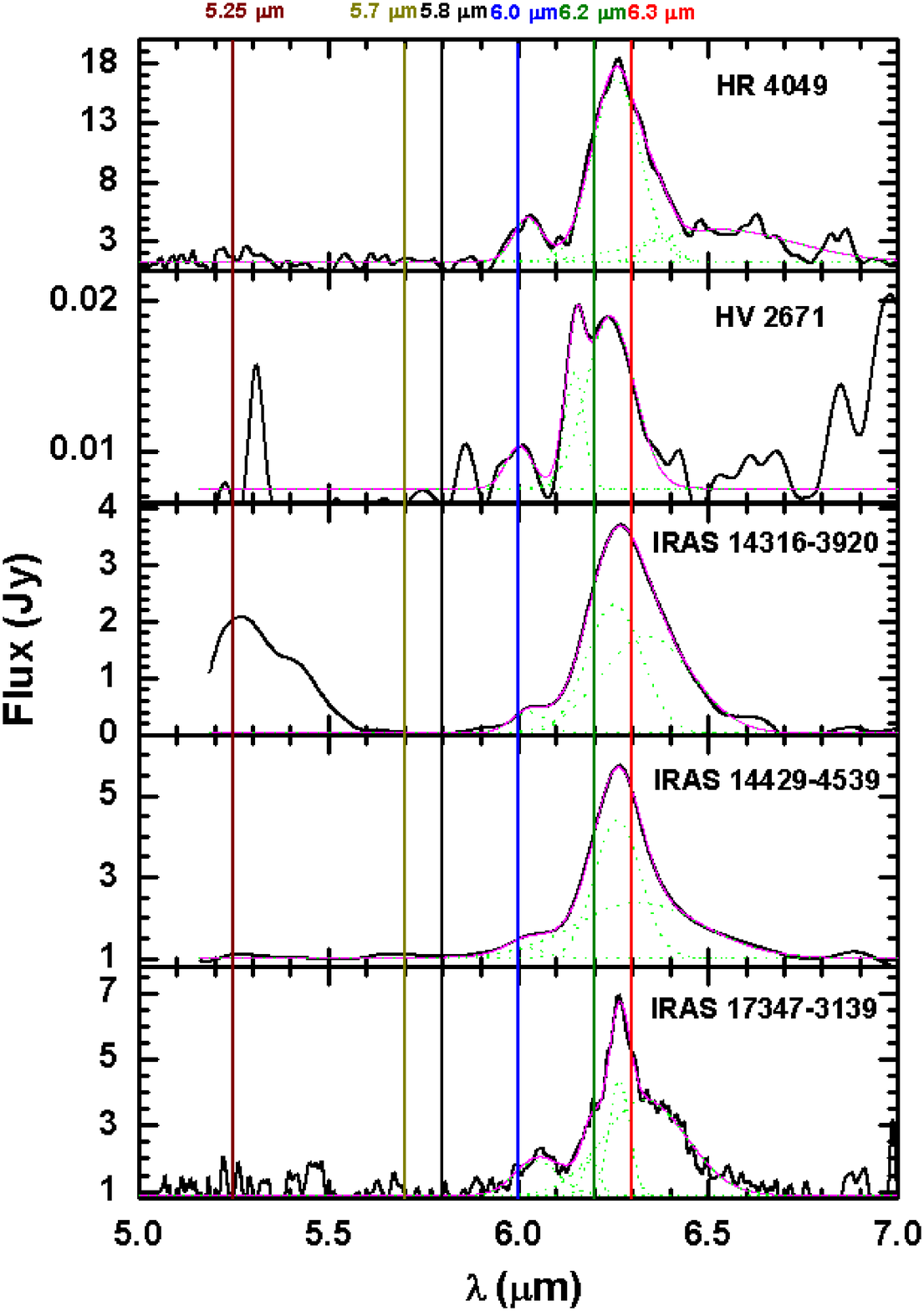}} \\
\end{tabular}
\end{center}
\end{figure*}

\begin{figure*}
\begin{center}
\begin{tabular}{c}
\resizebox{140mm}{!}{\includegraphics{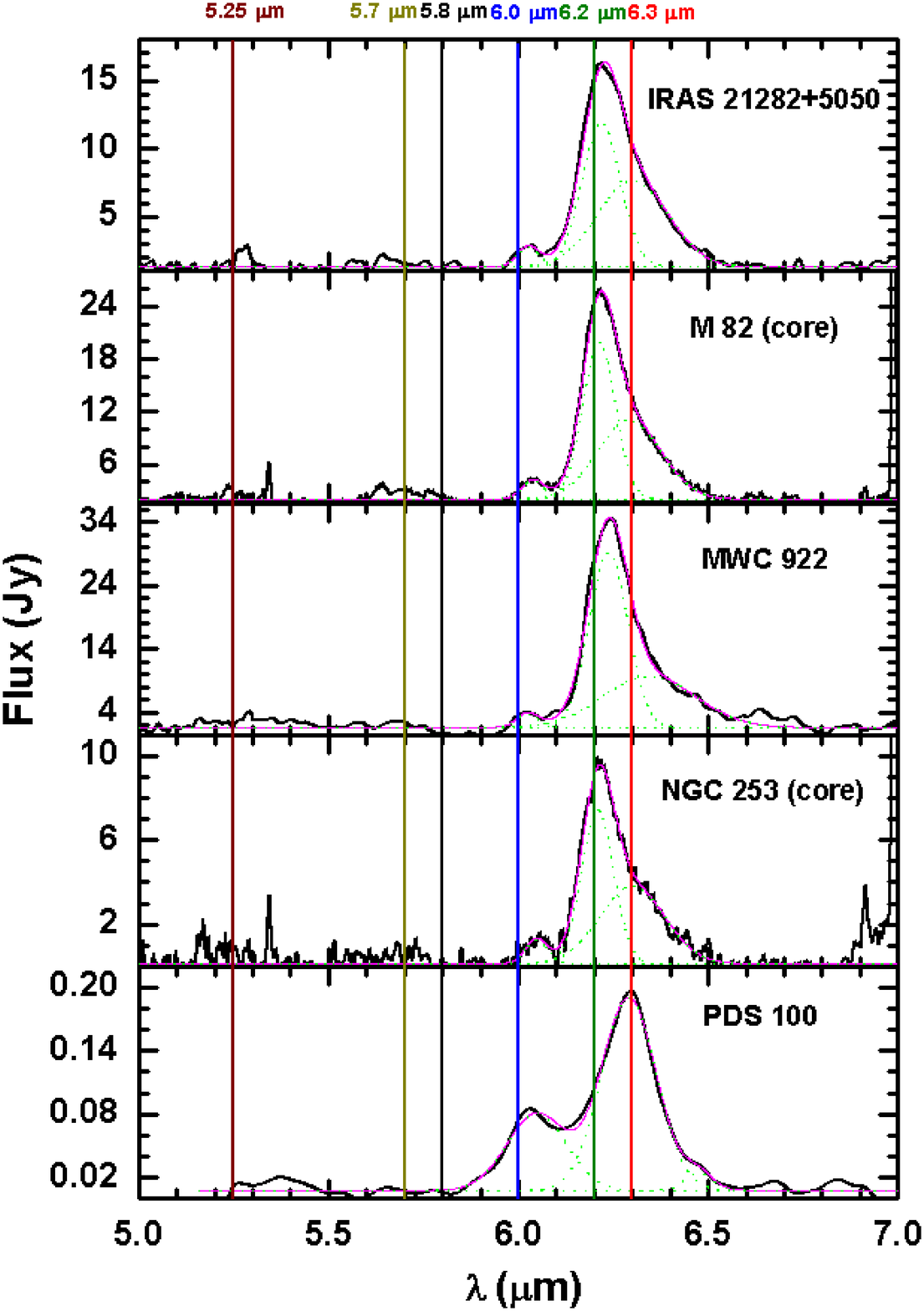}} \\
\end{tabular}
\end{center}
\end{figure*}

\begin{figure*}
\begin{center}
\begin{tabular}{c}
\resizebox{140mm}{!}{\includegraphics{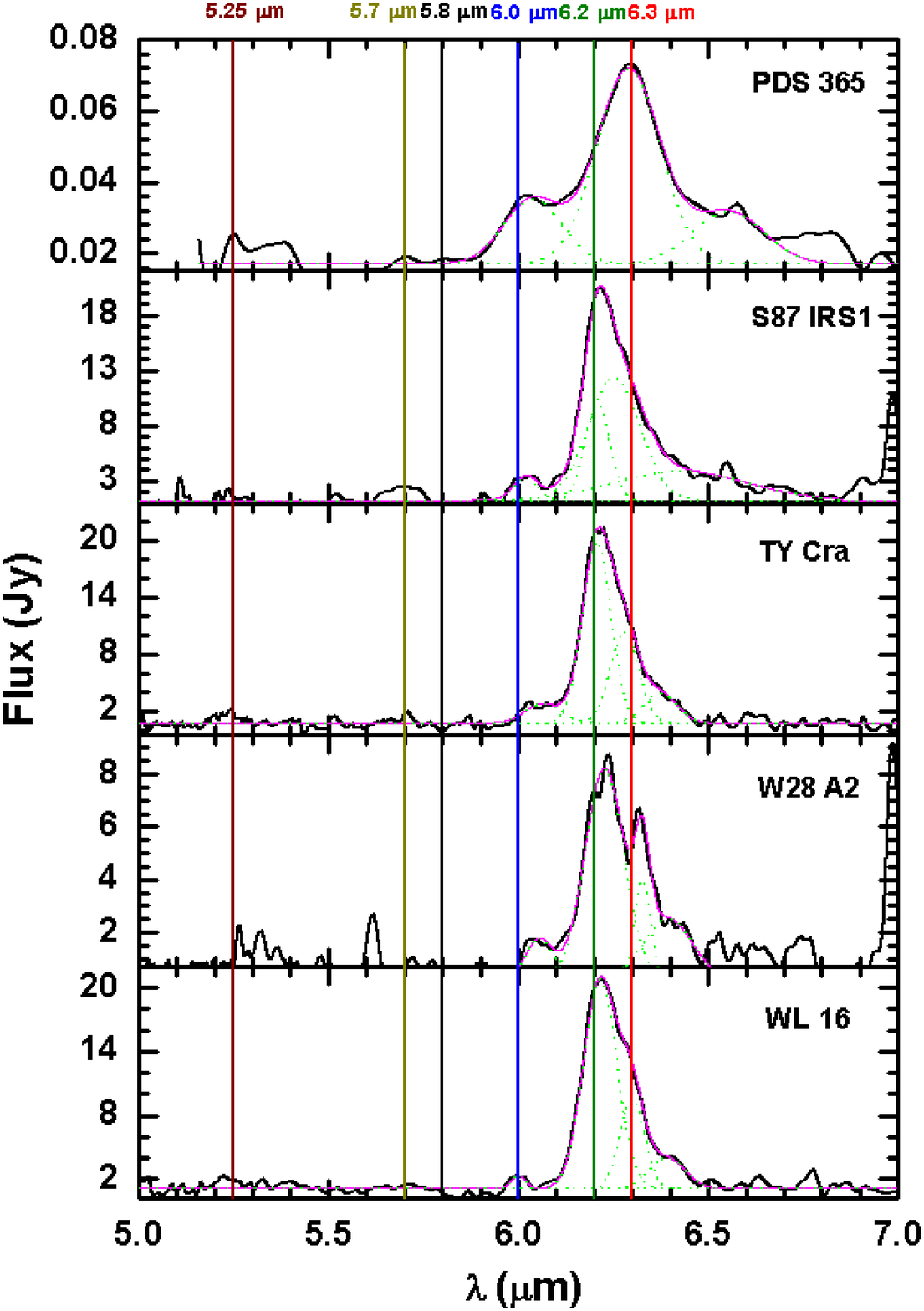}} \\
\end{tabular}
\end{center}
\end{figure*}
\clearpage

%
\clearpage

\clearpage

\clearpage

\clearpage

\clearpage
%
%
%
%
%
%
%
%
\clearpage
\clearpage

\clearpage


\clearpage


\begin{deluxetable}{llcccccccl}
\tabletypesize{\scriptsize} \rotate \tablecaption{Journal of Available Infrared Spectral Observations} \tablewidth{0pt}
\tablehead{\colhead{Source} & \colhead{R.A.} & \colhead{Decl.} & \colhead{Instrument} & \colhead{Class$^{a}$} & \colhead{Size$^{b}$} 
& \colhead{T$_{eff}$} & \colhead{G$_{0}^{c}$} & \colhead{n$_{e}$} & \colhead{Object Type} \\
\colhead{} & \colhead{(J2000.0)} & \colhead{(J2000.0)} & \colhead{} & \colhead{} & \colhead{($\arcsec$)} & \colhead{(10$^{3}$K)} 
& \colhead{} & \colhead{(cm$^{-3}$)} & \colhead{} \\
\hline
\multicolumn{10}{c}{ISO SWS Spectra} }
\startdata
NGC 253 (core) & 00:47:33.1 & -25:17:19.7 & TDT 24701422 & C$^{(1)}$ & 1260$\times$252$^{(13)}$ & ... & ... & ... & Galaxy \\
HD 44179 & 06:19:58.2 & -10:38:14.7 & TDT 70201801 & M$^{(2)}$ & 3.3$\times$5.9$^{(14)}$ & 9.52$^{(20)}$ & 5$\times$10$^{6}$$^{(3)}$ &  ... 
& PAGB \\
M 82$^{*}$ (core) & 09:55:52.4 & +69:40:46.9 & TDT 11600319 & C$^{(1)}$ & 304$\times$97$^{(15)}$ & ... & ... & ... & Galaxy \\
HR 4049 & 10:18:07.6 & -28:59:31.2 & TDT 17100101 & M$^{(2)}$ & 1.52$^{(16)}$ & 7.5$^{(21)}$ & ... & ... & PAGB \\
Circinus$^{*}$ (core) & 14:13:09.9 & -65:20:20.5 & TDT 07902231 & C$^{(3)}$ & 488$\times$200$^{(15)}$ & ... & ... & ... & Galaxy \\
WL 16$^{*}$ & 16:27:15.1 & -48:39:26.8 & TDT 48400535 & C$^{(4)}$ & ... & 9$^{(22)}$ & ... & ... & HAe \\
Hen 3-1333 & 17:09:00.9 & -56:54:48.1 & TDT 13602083 & M$^{(5)}$ & 5.6$\times$4.6$^{(14)}$ & 32$^{(12)}$ & 5$\times$10$^{6}$$^{(3)}$ 
& 1$\times$10$^{2}$$^{(30)}$ & YPN \\
IRAS 17347-3139 & 17:38:00.6 & -31:40:55.2 & TDT 87000939 & M$^{(6)}$ & 4.1$\times$3.5$^{(14)}$ & $>$26$^{(23)}$ & 8$\times$10$^{5}$$^{(3)}$ 
& 1$\times$10$^{6}$$^{(23)}$ & YPN \\
W28 A2$^{*}$ & 18:00:30.4 & -24:04:00.2 & TDT 12500843 & M$^{(7)}$ & 5$^{(17)}$ & 42$^{(17)}$ & ... & 1.1$\times$10$^{5}$$^{(17)}$ & CHII \\
MWC 922 & 18:21:16.1 & -13:01:25.6 & TDT 70301807 & C$^{(3)}$ & ... & 33$^{(24)}$ & 6$\times$10$^{6}$$^{(3)}$ & 1$\times$10$^{5}$$^{(31)}$ 
& Em star \\
TY Cra & 19:01:40.8 & -36:52:33.9 & TDT 34400603 & C$^{(8)}$ & ... & 12$^{(25)}$ & 6$\times$10$^{3}$$^{(29)}$ & ... & HAe/Be \\
S87 IRS1$^{*}$ & 19:46:20.1 & +24:35:29.4 & TDT 19200933 & C$^{(9)}$ & 7$\times$4$^{(18)}$ & 23$^{(18)}$ & 7$\times$10$^{6}$$^{(3)}$ 
& 1.4$\times$10$^{4}$$^{(18)}$& CHII \\
IRAS 21282+5050 & 21:29:58.4 & +51:03:59.8 & TDT 05602477 & C$^{(3)}$ & 6$\times$6$^{(19)}$ & 30$^{(19)}$ & 1$\times$10$^{5}$$^{(3)}$ 
& 2$\times$10$^{3}$$^{(19)}$ & YPN \\
\hline
\multicolumn{10}{c}{Spitzer IRS data} \\
\hline
HV 2671 & 05:33:48.9 & -70:13:23.4 & AOR key 18144512 & C$^{(5)}$ & ... & 20$^{(26)}$ & ... & ... & LMC RCB \\
HD 233517 & 08:22:46.7 & +53:04:49.2 & AOR key 3586048 & M$^{(10)}$ & ... & 4.48$^{(27)}$ & ... & ... & K Giant \\
DY Cen & 13:25:34.1 & -54:14:43.1 & AOR key 22274304 & C$^{(11)}$ & ... & 19.5$^{(26)}$ & ... & 4.5$\times$10$^{2}$$^{(26)}$ & RCB  \\
PDS 365 & 13:34:37.4 & -58:53:32.3 & AOR key 16263936 & M$^{(10)}$ & ... & 4.54$^{(10)}$ & ... & ... & K Giant \\
IRAS 14316-3920 & 14:34:49.4 & -39:33:19.2 & AOR key 17741056 & C$^{(12)}$ & U$^{(14)}$ & 6.75$^{(12)}$ & ... & ... & PAGB \\
IRAS 14429-4539 & 14:46:13.8 & -45:52:05.1 & AOR key 25453824 & C$^{(12)}$ & U$^{(14)}$ & 7$^{(28)}$ & ... & ... & PAGB \\
PDS 100 & 19:31:01.2 & +05:23:53.5 & AOR key 16263680 & M$^{(10)}$ & ... & 4.5$^{(10)}$ & ... & ... & K Giant \\
\enddata
\label{tab1}
\tablenotetext{{\it a}}{M: mixed-chemistry dust, C: carbon-rich dust.}
\tablenotetext{{\it b}}{U: unresolved.}
\tablenotetext{{\it c}}{G$_{0}$ is an estimate of ultraviolet flux density at the location where the emission originates from in 
units (1.6$\times$10$^{-6}$ W/m$^{2}$) of the average interstellar radiation field.}
\tablenotetext{{\it *}}{silicate absorption at 9.7 $\micron$ presented.}

\tablerefs{(1) Sturm et al. (2000); (2) van Winckel (2003); (3) Peeters et al. (2002); (4) DeVito \& Hayward (1998); 
(5) Clayton et al. (2011); (6) Jim\'enez-Esteban et al. (2006); (7) Volk \& Cohen (1989); (8) Boersma et al. (2008); 
(9) Hodge et al. (2004); (10) de la Reza et al. (2015); (11) Garc\'ia-Hern\'andez et al. (2011); (12) Szczerba et al. (2005).
(13) Uyama et al. (1984); (14) Lagadec et al. (2011); (15) Skrutskie et al. (2006); (16) Meixner et al. (1999); 
(17) Feldt et al. (1999); (18) Kurtz et al. (1994); (19) Likkel et al. (1994); (20) Men'shchikov et al. (2002); 
(21) Acke et al. (2013); (22) Ressler \& Barsony (2003); (23) de Gregorio-Monsalvo et al. (2004); (24) Rodr\'{i}guez et al. (2012); 
(25) Va\v{n}ko et al. (2013); (26) De Marco et al. (2002); (27) Sloan et al. (2007); (28) Cerrigone et al. (2011); (29) Hony et al. (2001); 
(30) Tylenda et al. (1991); (31) Rudy et al. (1992).}
\end{deluxetable}

\clearpage

\begin{deluxetable}{lcccc}
\tabletypesize{\footnotesize} \tablecaption{Measured Parameters of the 6.0 and 6.2 $\micron$ Features} \tablewidth{0pt}
\tablehead{\multicolumn{1}{c}{} & \multicolumn{2}{c}{6.0 $\micron$$^{(a)}$} & \multicolumn{1}{c}{6.2 $\micron$$^{(a)}$} &
\multicolumn{1}{c}{} \\
\cline{2-3}
\colhead{Source} & \colhead{$\lambda_{c}$} & \colhead{FHWM} & \colhead{$\lambda_{c}$} &
\colhead{$F_{\rm \scriptscriptstyle 6.0}$/$F_{\rm \scriptscriptstyle 6.2}$} \\
\colhead{} & \colhead{($\micron$)} & \colhead{($\micron$)} & \colhead{($\micron$)} & \colhead{}
}
\startdata
Circinus$^{(b)}$ (core) & 6.028$\pm$0.011 & 0.065$\pm$0.011 & 6.200$\pm$0.011 & 0.135$\pm$0.024 \\
DY Cen & 6.029$\pm$0.043 & 0.057$\pm$0.047 & 6.253$\pm$0.038 & 0.037$\pm$0.010 \\
HD 44179 & 6.021$\pm$0.011 & 0.070$\pm$0.011 & 6.264$\pm$0.011 &  0.092$\pm$0.018 \\
HD 233517 & 6.038$\pm$0.039 & 0.105$\pm$0.043 & 6.296$\pm$0.038 & 0.257$\pm$0.062 \\
Hen 3-1333 & 6.037$\pm$0.011 & 0.085$\pm$0.011 & 6.230$\pm$0.011 & 0.045$\pm$0.011 \\
HR 4049 & 6.023$\pm$0.011 & 0.073$\pm$0.011 & 6.263$\pm$0.011 & 0.083$\pm$0.016 \\
HV 2671 & 6.002$\pm$0.042 & 0.061$\pm$0.047 & 6.241$\pm$0.040 & 0.107$\pm$0.034 \\
IRAS 14316-3920 & 6.028$\pm$0.041 & 0.085$\pm$0.049 & 6.253$\pm$0.038 & 0.047$\pm$0.014 \\
IRAS 14429-4539 & 6.014$\pm$0.039 & 0.091$\pm$0.045 & 6.261$\pm$0.038 & 0.029$\pm$0.009 \\
IRAS 17347-3139 & 6.047$\pm$0.011 & 0.116$\pm$0.013 & 6.265$\pm$0.011 & 0.139$\pm$0.032 \\
IRAS 21282+5050 & 6.024$\pm$0.011 & 0.058$\pm$0.011 & 6.270$\pm$0.013 & 0.038$\pm$0.009 \\
M 82$^{(b)}$ (core) & 6.034$\pm$0.011 & 0.059$\pm$0.011 & 6.260$\pm$0.011 & 0.041$\pm$0.009 \\
MWC 922 & 6.024$\pm$0.011 & 0.056$\pm$0.013 & 6.237$\pm$0.012 & 0.025$\pm$0.007 \\
NGC 253$^{(b)}$ (core) & 6.042$\pm$0.011 & 0.057$\pm$0.011 & 6.257$\pm$0.012 & 0.053$\pm$0.013 \\
PDS 100 & 6.048$\pm$0.038 & 0.152$\pm$0.039 & 6.291$\pm$0.038 & 0.412$\pm$0.082 \\
PDS 365 & 6.038$\pm$0.039 & 0.148$\pm$0.042 & 6.290$\pm$0.038 & 0.233$\pm$0.061 \\
S87 IRS1 & 6.020$\pm$0.011 & 0.062$\pm$0.012 & 6.251$\pm$0.011 & 0.042$\pm$0.012 \\
TY Cra & 6.065$\pm$0.011 & 0.095$\pm$0.012 & 6.206$\pm$0.011 & 0.081$\pm$0.020 \\
W28 A2 & 6.055$\pm$0.011 & 0.071$\pm$0.012 & 6.226$\pm$0.011 & 0.084$\pm$0.019 \\
WL 16 & 6.000$\pm$0.011 & 0.038$\pm$0.011 & 6.215$\pm$0.011 & 0.018$\pm$0.005 \\
\enddata
\label{tab2}
\tablenotetext{{\it a}}{The overall uncertainties of $\lambda_{c}$ and FWHM are derived from typical systematic errors of instrument 
wavelength calibrations, spectral resolution estimations, and the errors of band wavelength determinations associated with our fitting 
procedure.}
\tablenotetext{{\it b}}{Redshifted correction, the redshift value is taken from the NASA/IPAC Extragalactic Database.}

\end{deluxetable}

\clearpage

\begin{deluxetable}{ccc}
\tabletypesize{\footnotesize} \tablecaption{The sample of molecules used in this study} \tablewidth{0pt}
\tablehead{\colhead{No.} & \colhead{Molecule} & \colhead{Formula}} 
\startdata
                & Molecules with experimental spectra & \\
\hline
        1       & Naphthalene & C$_{10}$H$_{8}$ \\
        2       & Anthracene  & C$_{14}$H$_{10}$  \\
        3   & phenanthrene & C$_{14}$H$_{10}$  \\
        4       & Pyrene & C$_{16}$H$_{10}$  \\
        5       & Tertracene & C$_{18}$H$_{12}$  \\
        6       & Chrysene & C$_{18}$H$_{12}$  \\
        7       & Benzo[a]anthracene & C$_{18}$H$_{12}$ \\
        8       & Benzo[ghi]perylene & C$_{22}$H$_{12}$  \\
        9       & Pentacene & C$_{22}$H$_{14}$ \\
        10      & Coronene & C$_{24}$H$_{12}$  \\
        11      &  UID 143  & C$_{48}$H$_{22}$ \\
        12      &  UID 149 & C$_{50}$H$_{22}$ \\
        13      & Ethane & C$_{2}$H$_{6}$ \\
        14      & 2-methylbutane & C$_{5}$H$_{12}$ \\
        15      & Ethylene & C$_{2}$H$_{4}$ \\
        16      & Propene & C$_{3}$H$_{6}$ \\
        17      & 1-butene & C$_{4}$H$_{8}$ \\
        18      & 1-pentene & C$_{5}$H$_{10}$ \\
        19      & 2-ethyl-3-methyl-1-butene & C$_{7}$H$_{14}$ \\
        20      & 1-3-butadiene & C$_{4}$H$_{6}$ \\
        21      & 1-4-pentadiene & C$_{5}$H$_{8}$ \\
        22      & 2,4-dimethyl-1,4-pentadiene &  C$_{7}$H$_{12}$ \\
        23      & acetylene & C$_{2}$H$_{2}$ \\
        24      & 1-butyne & C$_{4}$H$_{6}$ \\
        25      & Ethanol(Alcohol) & C$_{2}$H$_{5}$OH \\
        26      & Acrolien(Aldehyde) & C$_{2}$H$_{6}$O \\
        27      & propionaldehyde(Aldehyde) & C$_{3}$H$_{6}$O \\
        28      & Acetone(Ketone) & C$_{3}$H$_{6}$O \\
        29      & dimethyl ether(ether) & C$_{3}$H$_{6}$O \\
        30      & Acetic acid(carboxylic acid) & C$_{}$H$_{3}$COOH \\
        31      & Hexanoic acid(carboxylic acid) & C$_{5}$H$_{11}$COOH \\
        32      & propylene oxide(cyclic ether) & C$_{3}$H$_{6}$O \\
        33      & tetrahydrofuran (cyclic ether) & C$_{4}$H$_{8}$O \\
        34      & pheylacetaldehye & C$_{8}$H$_{8}$O \\
        35      & Allylbenzene & C$_{9}$H$_{10}$ \\
        36      & Allylnaphthalene  & C$_{13}$H$_{12}$ \\
\hline
\multicolumn{3}{c}{Group A} \\
\hline
        37      & Allylbenzene & C$_{9}$H$_{10}$ \\
        38      & Allylnaphthalene  & C$_{13}$H$_{12}$ \\
        39      & Allylanthracene & C$_{17}$H$_{14}$ \\
        40      & Allylphenanthrene & C$_{17}$H$_{14}$ \\
        41      & Allylpyrene & C$_{19}$H$_{14}$ \\
        42      & Allyltertracene & C$_{21}$H$_{16}$ \\
        43      & Allylchrysene & C$_{21}$H$_{16}$ \\
        44      & Allyltriphenylene & C$_{21}$H$_{16}$ \\
        45      & Allylbenzo[a]anthracene & C$_{21}$H$_{16}$ \\
        46      & Allylperylene & C$_{23}$H$_{16}$ \\
        47      & Allylbenzo[a]pyrene & C$_{23}$H$_{16}$ \\
        48      & Allylbenzo[e]pyrene &  C$_{23}$H$_{16}$ \\
        49      & Allylanthanthrene &  C$_{25}$H$_{16}$ \\
        50      & Allylbenzo[ghi]perylene &  C$_{25}$H$_{16}$ \\
        51      & Allylpentacene &  C$_{25}$H$_{18}$ \\
        52      & Allylcoronene &  C$_{17}$H$_{16}$ \\
        53      & Allyldibenzo[b,def]chrysene &  C$_{27}$H$_{18}$ \\
        54      & Allyldibenzo[cd,lm]perylene &  C$_{29}$H$_{18}$ \\
        55      & Allylhexacene &  C$_{29}$H$_{20}$ \\
        56      & AllylBisanthene  &  C$_{31}$H$_{18}$ \\
\hline
\multicolumn{3}{c}{Group B} \\
\hline
        57      & Phenylacetaldehyde  &  C$_{8}$H$_{8}$O \\
        58      & Naphthylacetaldehyde &  C$_{12}$H$_{10}$O \\
        59      & Anthrylacetaldehyde &  C$_{16}$H$_{12}$O \\
        60      & Phenanthrylacetaldehyde &  C$_{16}$H$_{12}$O \\
        61      & pyrene acetaldehyde &  C$_{18}$H$_{12}$O \\
        62      & tertracene acetaldehyde &  C$_{20}$H$_{14}$O \\
        63      & chrysene acetaldehyde &  C$_{20}$H$_{14}$O \\
        64      & triphenylene acetaldehyde &  C$_{20}$H$_{14}$O \\
        65      & benzo[a]anthracene acetaldehyde &  C$_{20}$H$_{14}$O \\
        66      & perylene acetaldehyde &  C$_{22}$H$_{14}$O \\
        67      & benzo[a]pyrene acetaldehyde &  C$_{22}$H$_{14}$O \\
        68      & benzo[e]pyrene acetaldehyde &  C$_{22}$H$_{14}$O \\
        69      & anthanthrene acetaldehyde &  C$_{}$H$_{}$O \\
        70      & benzo[ghi]perylene acetaldehyde &  C$_{24}$H$_{14}$O \\
        71      & pentacene acetaldehyde &  C$_{24}$H$_{16}$O \\
        72      & coronene acetaldehyde &  C$_{26}$H$_{14}$O \\
        73      & dibenzo[b,def]chrysene acetaldehyde &  C$_{26}$H$_{16}$O \\
        74      & dibenzo[cd,lm]perylene acetaldehyde &  C$_{28}$H$_{16}$O \\
        75      & hexacene acetaldehyde &  C$_{28}$H$_{18}$O \\
        76      & Bisanthene acetaldehyde &  C$_{30}$H$_{16}$O \\
\hline
	& PAH cores& \\
\hline
77  &Benzene &C$_{6}$H$_{6}$  \\ 
78  &Naphthalene & C$_{10}$H$_{8}$  \\ 
79 &Anthracene & C$_{14}$H$_{10}$  \\ 
80  &phenanthrene & C$_{14}$H$_{10}$  \\ 
81  &Pyrene & C$_{16}$H$_{10}$  \\ 
82  &Tertracene & C$_{18}$H$_{12}$  \\ 
83  &Chrysene & C$_{18}$H$_{12}$  \\
84  &Triphenylene & C$_{18}$H$_{12}$  \\ 
85  &Benzo[a]anthracene & C$_{18}$H$_{12}$  \\ 
86 &Perylene &C$_{20}$H$_{12}$  \\ 
87 &Benzo[a]pyrene &C$_{20}$H$_{12}$  \\
88 &Benzo[e]pyrene &C$_{20}$H$_{12}$ \\ 
89 &Anthanthrene &C$_{22}$H$_{12}$ \\ 
90 &Benzo[ghi]perylene &C$_{22}$H$_{12}$ \\
91 &Pentacene &C$_{22}$H$_{14}$ \\
92 &Coronene &C$_{24}$H$_{12}$ \\
93 &Dibenzo[b,def]chrysene &C$_{24}$H$_{14}$ \\ 
94 &Dibenzo[cd,lm]perylene &C$_{26}$H$_{14}$ \\ 
95 &Hexacene &C$_{26}$H$_{16}$ \\
96 &Bisanthene &C$_{28}$H$_{14}$ \\ \enddata
\label{tab3}
\end{deluxetable}

\clearpage

\begin{deluxetable}{ccc}
\tabletypesize{\footnotesize} 
\tablecaption{Peak Positions of the Simulated Bands around 6.0 and 6.2 $\micron$ for Group A molecules} \tablewidth{0pt}
\tablehead{\colhead{No.} & \colhead{6.0 $\micron$} & \colhead{6.2 $\micron$}}
\startdata
37      & 6.00 & 6.19  \\
38      & 6.01 & 6.24 \\
39      & --$^a$ & 6.14 \\
40      & 6.00 & 6.24 \\
41      & 6.01 & 6.24 \\
42      & 6.06 & 6.40 \\
43      & --$^a$ & 6.14  \\
44      & 6.02 & 6.15 \\
45      & --$^a$ & 6.14  \\
46      & --$^a$ & 6.22 \\
47      & --$^a$ & 6.13 \\
48      & --$^a$ & 6.21 \\
49      & 6.12 & 6.23 \\
50      & --$^a$ & 6.21 \\
51      & 6.05 &  \\
52      & --$^a$ & 6.17 \\
53      & 6.11 &  \\
54      & --$^a$ & 6.30 \\
55      & 6.05 &  \\
56      &  --$^a$  & 6.15 \\
\enddata
\tablenotetext{{\it a}}{under the shoulder of the 6.2 $\mu$m feature.}
\label{tab4}
\end{deluxetable}

\end{CJK*}

\end{document}